\documentclass[aps,prb,reprint,longbibliography,noeprint,superscriptaddress]{revtex4-2}
\usepackage{amsfonts,amssymb,amsmath,bm,graphicx,xcolor}
\usepackage[colorlinks=true,citecolor=blue,linkcolor=blue,urlcolor=blue]{hyperref}
\allowdisplaybreaks[4]
\newcommand{\cc}{\mathrm{c.c.}}
\newcommand{\hb}[1]{\hat{\bm{#1}}}
\DeclareMathOperator{\tr}{tr}
\begin{document}
\title{Intrinsic spin accumulation in the magnetic spin Hall effect}
\author{Atsuo Shitade}
\affiliation{Institute of Scientific and Industrial Research, The University of Osaka, Ibaraki, Osaka 567-0047, Japan}
\date{\today}
\begin{abstract}
  The magnetic spin Hall effect is a time-reversal-odd phenomenon in which spin current is induced by the charge current.
  In the presence of a spin-orbit coupling and/or noncolinear magnetism, however, spin current is not uniquely defined.
  Instead, we study an intrinsic response of spin to an electric field gradient
  that describes the spin accumulation at the boundaries of magnetic systems.
  We derive a generic formula expressed by Bloch wave functions
  and apply it to a minimal model for representative altermagnets, RuO$_{2}$ and MnF$_{2}$.
  Our results show that the intrinsic spin accumulation can be nonzero in magnetic insulators
  in sharp contrast to the magnetic spin Hall conductivity.
\end{abstract}
\maketitle
\section{Introduction} \label{sec:introduction}
Spin Hall (SH) effect has attracted scientific and technological interests
as a means of electrical generation of spin current~\cite{RevModPhys.87.1213},
which turns into the spin accumulation at the boundaries of the system~\cite{Dyakonov1971}.
This phenomenon is allowed even in nonmagnetic systems owing to spin-orbit coupling (SOC).
The research area started from the theoretical rediscoveries~\cite{Murakami1348,PhysRevLett.92.126603}
and experimental observations in semiconductors~\cite{PhysRevLett.94.047204,Kato1910}.

Spin transport
in ferromagnets~\cite{PhysRevApplied.3.044001,PhysRevB.106.024410,PhysRevB.109.224401,10.1063/5.0255152,Iihama2018,PhysRevApplied.9.064033,PhysRevB.96.220408,Humphries2017}
and antiferromagnets (AFMs)~\cite{RevModPhys.90.015005,Kimata2019,PhysRevLett.124.087204,Nan2020,PhysRevLett.128.197202,PhysRevLett.129.137201,Bose2022,Chen2021,Naka2019,PhysRevResearch.2.023065,PhysRevLett.126.127701}
is of recent interest.
Since a magnetic order breaks not only the time-reversal symmetry but also some spatial symmetries,
the unconventional components forbidden in the paramagnetic phase may be allowed~\cite{PhysRevB.92.155138,PhysRevLett.119.187204}.
In addition, the time-reversal odd parts imply that the spin direction can be controlled by the magnetization reversal.
This phenomenon is referred to as the magnetic SH effect~\cite{Kimata2019} and potentially applied to new spintronics devices.
The SH conductivity, namely, the linear response of spin current to an electric field,
has been widely evaluated as an indicator of the conventional~\cite{PhysRevLett.94.226601,PhysRevLett.95.156601}
and magnetic SH effects~\cite{PhysRevB.106.024410,PhysRevB.109.224401,10.1063/5.0255152,PhysRevResearch.2.023065,PhysRevLett.126.127701,PhysRevB.92.155138,PhysRevLett.119.187204}.

In the presence of SOC and/or noncolinear magnetism, spin is not conserved, and spin current cannot be uniquely defined.
One choice is the conventional spin current $\hat{J}_{sa}^{\phantom{sa} i} = \{\hat{s}_{a}, \hat{v}^{i}\} / 2$,
where $\hat{s}_{a} = (\hbar / 2) \sigma_{a}$ and $\hat{v}^{i}$ are the spin and velocity operators, respectively,
with the Pauli matrix $\sigma_{a}$.
This choice suffers from some critical problems such as
the nonzero equilibrium expectation value in noncentrosymmetric systems~\cite{PhysRevB.68.241315} and the absence of a conjugate force.
Another choice called the conserved spin current,
which consists of the conventional one and spin torque dipole moment~\cite{PhysRevLett.96.076604,PhysRevB.104.L241411},
can overcome these problems.
However, it remains unsolved what definition of spin current is proper and experimentally observed.

An alternative indicator is the spin accumulation coefficient (SAC), namely, 
the linear response of spin to an electric field gradient~\cite{PhysRevB.98.174422,PhysRevB.105.L201202,PhysRevB.106.045203}.
This quantity characterizes the spin accumulation at the boundaries of the system
but can be evaluated as a bulk property using Bloch wave functions.
Recently, an \textit{ab initio} computational scheme has been implemented~\cite{Shitade2025}
in an open-source software Wannier90~\cite{Pizzi_2020}.
Compared with the SH conductivity, this quantity has the advantages that spin is well defined
and that the spin accumulation has been experimentally observed~\cite{Kato1910,PhysRevLett.94.047204}.
However, the above theories~\cite{PhysRevB.98.174422,PhysRevB.105.L201202,PhysRevB.106.045203} focused on the extrinsic terms only.
In the absence of time-reversal symmetry, an intrinsic term is allowed.

In this paper, we study the intrinsic contribution to the SAC for the magnetic SH effect.
While the magnetic SH conductivity vanishes in insulators, this contribution can be nonzero.
We rely on the quantum-mechanical linear response theory to derive a generic formula expressed by Bloch wave functions.
We apply this formula to a minimal model for altermagnets, RuO$_{2}$ and MnF$_{2}$~\cite{PhysRevB.110.144412,PhysRevLett.134.096703}.
The former is metallic, and its magnetic SH effect has been experimentally studied~\cite{PhysRevLett.128.197202,PhysRevLett.129.137201,Bose2022}.
The latter is insulating and hence a good example to demonstrate the difference between the magnetic SH conductivity and intrinsic SAC.
We also discuss a relation to the spin magnetic octupole moment recently formulated~\cite{PhysRevLett.131.106701,PhysRevB.112.134412,2504.21431}.

\section{Terminology} \label{sec:terminology}
Throughout this paper, we distinguish the conventional and magnetic SH effects as follows.
In terms of spin current, these phenomena are characterized by the responses to an electric field and the charge current, respectively:
\begin{subequations}\begin{align}
  \langle \Delta \hat{J}_{sa}^{\phantom{sa} i} \rangle
  = & \sigma_{sa}^{(\mathrm{II}) ij} E_{j}, \label{eq:conventional_shc} \\
  \langle \Delta \hat{J}_{sa}^{\phantom{sa} i} \rangle
  = & \theta_{sa \phantom{i} k}^{(\mathrm{I}) i} \langle \Delta \hat{J}^{k} \rangle. \label{eq:magnetic_shc}
\end{align}\label{eq:shc}\end{subequations}
Here, $\hat{J}_{sa}^{\phantom{sa} i}$ includes, but is not limited to, the conventional spin current.
In Eq.~\eqref{eq:magnetic_shc}, the charge current is induced by the electric field as
$\langle \Delta \hat{J}^{k} \rangle = \sigma^{kj} E_{j}$.
Directly computed with the Kubo formula is not $\theta_{sa \phantom{i} k}^{(\mathrm{I}) i}$
but $\sigma_{sa}^{(\mathrm{I}) ij} = \theta_{sa \phantom{i} k}^{(\mathrm{I}) i} \sigma^{kj}$.
Nonetheless, Eq.~\eqref{eq:shc} is fundamental and useful for determining which components are allowed by symmetry.

\begin{table*}
  \centering
  \caption{%
    Allowed or forbidden responses of spin current and spin in the conventional and magnetic SH effects.
    Our main results are highlighted with $\ast$.%
  } \label{tab:she}
  \begin{tabular}{cccccc} \hline \hline
    & & \multicolumn{2}{c}{Conventional SH effect} & \multicolumn{2}{c}{Magnetic SH effect} \\ \cline{3-6}
    & & Spin current & Spin & Spin current & Spin \\ \hline
    Nonmagnetic & metals & $\checkmark$ & $\checkmark$ & & \\
    & insulators & $\checkmark$ & & & \\
    Magnetic & metals & $\checkmark$ & $\checkmark$ & $\checkmark$ & $\checkmark^{\ast}$ \\
    & insulators & $\checkmark$ & & & $\checkmark^{\ast}$ \\ \hline \hline
  \end{tabular}
\end{table*}
The conventional SH effect characterized by Eq.~\eqref{eq:conventional_shc} is time-reversal even
and allowed in the presence and absence of the time-reversal symmetry.
$\sigma_{sa}^{(\mathrm{II}) ij}$ is expressed by the intrinsic contribution to the Kubo formula (interband Fermi-sea term)
and can be nonzero in metals and insulators, as summarized in the first column of Table~\ref{tab:she}.
Note that $\sigma_{sa}^{(\mathrm{II}) ij}$ is not always antisymmetric with respect to $i$ and $j$,
though we call the phenomenon the spin \textit{Hall} effect.
For instance, trigonal tellurium, which belongs to the $D_{3}$ point group, has nonzero planar components,
$\sigma_{sx}^{(\mathrm{II}) xx} = -\sigma_{sx}^{(\mathrm{II}) yy}
= -\sigma_{sy}^{(\mathrm{II}) xy} = -\sigma_{sy}^{(\mathrm{II}) yx}$~\cite{Shitade2025}.
The spin anomalous Hall effect in ferromagnets~\cite{PhysRevApplied.3.044001,Iihama2018,PhysRevApplied.9.064033,PhysRevB.96.220408}
is categorized into this phenomenon.

The magnetic SH effect characterized by Eq.~\eqref{eq:magnetic_shc} is time-reversal odd
and allowed only when the time-reversal symmetry is broken.
$\sigma_{sa}^{(\mathrm{I}) ij}$ is expressed by the extrinsic contribution to the Kubo formula
(intraband Fermi-surface term)~\cite{PhysRevB.106.024410,PhysRevB.109.224401,10.1063/5.0255152,PhysRevResearch.2.023065,PhysRevLett.126.127701,PhysRevB.92.155138,PhysRevLett.119.187204}
and vanishes in insulators, as summarized in the third column of Table~\ref{tab:she}.
The spin splitter effect~\cite{PhysRevLett.128.197202,PhysRevLett.129.137201,Bose2022,PhysRevLett.126.127701},
for which $\sigma_{sa}^{(\mathrm{I}) ij}$ is symmetric with respect to $i$ and $j$, is categorized into this phenomenon,
while Ref.~\cite{PhysRevResearch.2.023065} focused on the antisymmetric components.
The longitudinal conductivity $\sigma_{sa}^{(\mathrm{I}) ii}$ in ferromagnets is also categorized into this phenomenon.

In terms of spin, the conventional and magnetic SH effects are characterized by
the responses to the gradients of the charge current and an electric field, respectively:
\begin{subequations}\begin{align}
  \langle \Delta \hat{s}_{a} \rangle
  = & \gamma_{sa \phantom{i} k}^{(\mathrm{I}) i} \partial_{x^{i}} \langle \Delta \hat{J}^{k} \rangle, \label{eq:conventional_sac} \\
  \langle \Delta \hat{s}_{a} \rangle
  = & g_{sa}^{(\mathrm{II}) ij} \partial_{x^{i}} E_{j}. \label{eq:magnetic_sac}
\end{align}\label{eq:sac}\end{subequations}
The symmetry-allowed structures of Eqs.~\eqref{eq:conventional_sac} and \eqref{eq:magnetic_sac}
coincide with those of Eqs.~\eqref{eq:conventional_shc} and \eqref{eq:magnetic_shc}.
In Eq.~\eqref{eq:conventional_sac}, directly computed with the Kubo formula is not $\gamma_{sa \phantom{i} k}^{(\mathrm{I}) i}$
but $g_{sa}^{(\mathrm{I}) ij} = \gamma_{sa \phantom{i} k}^{(\mathrm{I}) i} \sigma^{kj}$.

$g_{sa}^{(\mathrm{I}) ij}$ for the conventional SH effect
is expressed by one of the extrinsic contributions to the Kubo formula~\cite{PhysRevB.105.L201202}.
Hence, this coefficient vanishes in insulators, as summarized in the second column of Table~\ref{tab:she},
which is in a sharp contrast to $\sigma_{sa}^{(\mathrm{II}) ij}$.
This fact clearly indicates that the spin current induced by an electric field is not related to the spin accumulation.

$g_{sa}^{(\mathrm{II}) ij}$ for the magnetic SH effect consists of the intrinsic contribution derived in this paper
and the other extrinsic contribution to the Kubo formula~\cite{PhysRevB.105.L201202}.
The former can be nonzero in metals and insulators in contrast to $\sigma_{sa}^{(\mathrm{I}) ij}$,
as summarized in the fourth column of Table~\ref{tab:she}.
Thus, if the magnetic SH effect can be observed in insulators,
it should be attributed to the field-induced spin polarization, rather than spin current.

\section{Derivation} \label{sec:derivation}
In this section, we sketch the derivation of the intrinsic contribution to the SAC,
which is allowed only when the time-reversal symmetry is broken.
See details for Appendix~\ref{app:spin_current}.
We begin with the spin--charge-current correlation function,
\begin{align}
  \chi_{\hat{s}_{a} \hat{J}^{j}}^{\mathrm{R}}(\Omega, \bm{Q})
  = & -\frac{q}{\hbar} \sum_{nm} \int \frac{d^{d} k}{(2 \pi)^{d}} \notag \\
  & \times \langle u_{n} (\bm{k}_{-}) | \hat{s}_{a} | u_{m}(\bm{k}_{+}) \rangle \notag \\
  & \times \langle u_{m}(\bm{k}_{+}) | \hbar \hat{v}^{j}(\bm{k}; \bm{Q}) | u_{n}(\bm{k}_{-}) \rangle \notag \\
  & \times \frac{
    f(\epsilon_{n}(\bm{k}_{-})) - f(\epsilon_{m}(\bm{k}_{+}))
  }{
    \hbar \Omega + \epsilon_{n}(\bm{k}_{-}) - \epsilon_{m}(\bm{k}_{+}) + i \eta
  } \notag \\
  = & \chi_{\hat{s}_{a} \hat{J}^{j}}^{\mathrm{R}}(0, \bm{Q})
  + (i \Omega) \alpha_{\hat{s}_{a} \hat{J}^{j}}^{\mathrm{R}}(\Omega, \bm{Q}), \label{eq:spin_current_copy}
\end{align}
which describes the linear response of spin to a vector potential,
$\langle \Delta \hat{s}_{a} \rangle(\Omega, \bm{Q}) = \chi_{\hat{s}_{a} \hat{J}^{j}}^{\mathrm{R}}(\Omega, \bm{Q}) A_{j}(\Omega, \bm{Q})$.
Here, $q$ is the electron charge, $d$ is the spatial dimension, $\eta \rightarrow +0$ is the convergence factor,
$\bm{k}_{\pm} = \bm{k} \pm \bm{Q}/2$, and $f(\epsilon) = [e^{(\epsilon - \mu)/T} + 1]^{-1}$ is the Fermi distribution function.
$\hat{s}_{a}$ and
$\hat{v}^{j}(\bm{k}; \bm{Q}) = [\hat{v}^{j}(\bm{k}_{+}) + \hat{v}^{j}(\bm{k}_{-})] / 2$
are the spin and velocity operators, respectively.
Since we are interested in the response to an electric field gradient,
we expand $\alpha_{\hat{s}_{a} \hat{J}^{j}}^{\mathrm{R}}(\Omega, \bm{Q})$ up to the first order with respect to $\bm{Q}$.

Let us comment on the gauge invariance regarding Bloch wave functions.
Since the Kubo formula~\eqref{eq:spin_current_copy} is gauge-invariant,
it is better to decompose it into gauge-covariant objects.
For this reason, we rewrite the matrix elements as
\begin{align}
  & \langle u_{n} (\bm{k}_{-}) | \hat{s}_{a} | u_{m}(\bm{k}_{+}) \rangle
  \langle u_{m}(\bm{k}_{+}) | \hbar \hat{v}^{j}(\bm{k}; \bm{Q}) | u_{n}(\bm{k}_{-}) \rangle \notag \\
  = & \tr [\hat{s}_{a} | u_{m}(\bm{k}_{+}) \rangle \langle u_{m}(\bm{k}_{+}) |
  \hbar \hat{v}^{j}(\bm{k}; \bm{Q}) | u_{n}(\bm{k}_{-}) \rangle \langle u_{n} (\bm{k}_{-}) |], \label{eq:spin_current_3_copy}
\end{align}
and expand a gauge-covariant object $| u_{n}(\bm{k}_{\pm}) \rangle \langle u_{n}(\bm{k}_{\pm}) |$ with respect to $\bm{Q}$.
Hereafter, we omit the argument of $\bm{k}$ for simplicity.
Furthermore, we introduce the covariant derivative as
$| D_{k_{i}} u_{n} \rangle  = | \partial_{k_{i}} u_{n} \rangle + i | u_{n} \rangle a_{n}^{i}$,
where $a_{n}^{i} = i \langle u_{n} | \partial_{k_{i}} u_{n} \rangle$ is the intraband Berry connection.
We can easily show
$\partial_{k_{i}} (| u_{n} \rangle \langle u_{n} |) = | D_{k_{i}} u_{n} \rangle \langle u_{n} | + | u_{n} \rangle \langle D_{k_{i}} u_{n} |$.

In order to evaluate the spin-related matrix elements, we consider $\tilde{H} = \hat{H} + \hb{s} \cdot \bm{B}$
and take the $\bm{B} \rightarrow 0$ limit at the end of the calculation.
We introduce the interband spin magnetic quadrupole
and orbital magnetic moments~\cite{PhysRevB.98.060402,PhysRevB.103.045401,PhysRevB.107.094106} as
\begin{widetext}
  \begin{subequations}\begin{align}
    s_{mna}^{\phantom{mna} j}
    = & \frac{i}{2} \sum_{l (\not= n)} \frac{
      (s_{na} \delta_{ml} + \langle u_{m} | \hat{s}_{a} | u_{l} \rangle)
      \langle u_{l} | \hbar \hat{v}^{j} | u_{n} \rangle
      - (\partial_{k_{j}} \epsilon_{n} \delta_{ml} + \langle u_{m} | \hbar \hat{v}^{j} | u_{l} \rangle)
      \langle u_{l} | \hat{s}_{a} | u_{n} \rangle
    }{
      \epsilon_{n} - \epsilon_{l}
    }, \label{eq:spin_magnetic_quadrupole_copy} \\
    \epsilon^{ija} m_{mna}
    = & -\frac{i}{2} \sum_{l (\not= n)} \frac{
      (\partial_{k_{i}} \epsilon_{n} \delta_{ml} + \langle u_{m} | \hbar \hat{v}^{i} | u_{l} \rangle)
      \langle u_{l} | \hbar \hat{v}^{j} | u_{n} \rangle
      - (i \leftrightarrow j)
    }{
      \epsilon_{n} - \epsilon_{l}
    }, \label{eq:orbital_magnetic_copy}
  \end{align}\label{eq:magnetic_multipole_copy}\end{subequations}
  with $s_{na} = \langle u_{n} | \hat{s}_{a} | u_{n} \rangle$.
  Note that the sign of Eq.~\eqref{eq:orbital_magnetic_copy} is opposite to
  that in Refs.~\cite{PhysRevB.98.060402,PhysRevB.103.045401,PhysRevB.107.094106}.
  These quantities are non-Hermitian, but their diagonal elements are real
  and coincide with the intraband spin magnetic quadrupole~\cite{PhysRevB.99.024404,PhysRevB.105.L201202}
  and orbital magnetic moments, respectively.

  After lengthy calculation, we arrive at the first-order coefficients with respect to $\bm{Q}$ for $\Omega \rightarrow 0$,
  $\langle \hat{s}_{a} \rangle(0, \bm{Q})
  = \lim_{\Omega \rightarrow 0} g_{sa}^{\phantom{sa} ij} (i Q_{i}) (i \Omega) A_{j}(\Omega, \bm{Q})$.
  The conventional~\cite{PhysRevB.105.L201202} and magnetic parts are expressed by
  \begin{subequations}\begin{align}
    g_{sa}^{(\mathrm{I}) ij}
    = & -\frac{q}{\eta} \sum_{n} \int \frac{d^{d} k}{(2 \pi)^{d}}
    (s_{na}^{\phantom{na} i} \partial_{k_{j}} \epsilon_{n} - s_{na} \epsilon^{ijb} m_{nb}) [-f^{\prime}(\epsilon_{n})],
    \label{eq:spin_accumulation_1_copy} \\
    g_{sa}^{(\mathrm{II}) ij}
    = & -q \sum_{n \not= m} \int \frac{d^{d} k}{(2 \pi)^{d}} \left[
      \frac{
        i \langle u_{n} | \hbar \hat{v}^{j} | u_{m} \rangle s_{mna}^{\phantom{mna} i}
        + i \langle u_{n} | \hat{s}_{a} | u_{m} \rangle \epsilon^{ijb} m_{mnb} + \cc
      }{
        (\epsilon_{n} - \epsilon_{m})^{2}
      } f(\epsilon_{n})
    \right. \notag \\
    & \left.
      - \frac{1}{2} \frac{
        \langle u_{n} | \hat{s}_{a} | u_{m} \rangle \langle u_{m} | \hbar \hat{v}^{j} | u_{n} \rangle + \cc
      }{
        (\epsilon_{n} - \epsilon_{m})^{2}
      } \partial_{k_{i}} \epsilon_{n} f^{\prime}(\epsilon_{n})
    \right]
    - \frac{q}{\eta^{2}} \sum_{n} \int \frac{d^{d} k}{(2 \pi)^{d}}
    s_{na} \partial_{k_{i}} \epsilon_{n} \partial_{k_{j}} \epsilon_{n} [-f^{\prime}(\epsilon_{n})]. \label{eq:spin_accumulation_2_copy}
  \end{align}\label{eq:spin_accumulation_copy}\end{subequations}
\end{widetext}
The conventional part~\eqref{eq:spin_accumulation_1_copy} is purely extrinsic,
while the magnetic part~\eqref{eq:spin_accumulation_2_copy} consists of two terms;
the first term is intrinsic, which is the main result of this paper, and the second term is extrinsic.
The latter is similar to
the magnetic SH conductivity~\cite{PhysRevB.106.024410,PhysRevB.109.224401,10.1063/5.0255152,PhysRevResearch.2.023065,PhysRevLett.126.127701,PhysRevB.92.155138,PhysRevLett.119.187204}
if $s_{n} \partial_{k_{i}} \epsilon_{n}$ is replaced by $\langle u_{n} | \hbar \hat{J}_{sa}^{\phantom{sa} i} | u_{n} \rangle$,
but symmetric with respect to $i$ and $j$
though the antisymmetric components are allowed in the magnetic SH conductivity~\cite{PhysRevResearch.2.023065}.

In the collinear case without SOC, one, say $z$, component of spin is a good quantum number,
and Eq.~\eqref{eq:spin_accumulation_2_copy} is reduced to
\begin{align}
  g_{sz}^{(\mathrm{II}) ij}
  = & q \sum_{n^{\prime} \sigma} \int \frac{d^{d} k}{(2 \pi)^{d}}
  \frac{\hbar}{2} \sigma X_{n^{\prime} \sigma}^{\phantom{n^{\prime} \sigma} ij}
  f(\epsilon_{n^{\prime} \sigma}) \notag \\
  & - \frac{q}{\eta^{2}} \sum_{n^{\prime} \sigma} \int \frac{d^{d} k}{(2 \pi)^{d}}
  \frac{\hbar}{2} \sigma \partial_{k_{i}} \epsilon_{n^{\prime} \sigma} \partial_{k_{j}} \epsilon_{n^{\prime} \sigma}
  [-f^{\prime}(\epsilon_{n^{\prime} \sigma})], \label{eq:spin_accumulation_wo_soc}
\end{align}
with $n = (n^{\prime}, \sigma)$, where
\begin{equation}
  X_{n}^{\phantom{n} ij}
  = -\sum_{m (\not= n)} \frac{
    \langle u_{n} | \hbar \hat{v}^{i} | u_{m} \rangle \langle  u_{m} | \hbar \hat{v}^{j} |  u_{n} \rangle + \cc
  }{
    (\epsilon_{n} - \epsilon_{m})^{3}
  }, \label{eq:electric_susceptibility}
\end{equation}
appears in the formulas of the electric susceptibility and electric quadrupole moment~\cite{PhysRevB.102.235149}.
Equation~\eqref{eq:spin_accumulation_wo_soc} is symmetric with respect to $i$ and $j$.

\section{Application to Altermagnets} \label{sec:altermagnets}
In this section, we apply our formula~\eqref{eq:spin_accumulation_copy} to two representative altermagnets, RuO$_{2}$ and MnF$_{2}$.
These compounds are collinear AFMs on the rutile structure.
The magnetic point group $4^{\prime}/mmm^{\prime}$ allows the forms of
\begin{subequations}\begin{align}
  g_{sx}^{(\mathrm{I})}
  = & \begin{bmatrix}
    0 & 0 & 0 \\
    0 & 0 & g_{sx}^{(\mathrm{I}) yz} \\
    0 & g_{sx}^{(\mathrm{I}) zy} & 0
  \end{bmatrix}, \label{eq:spin_accumulation_1_altermagnets_x} \\
  g_{sy}^{(\mathrm{I})}
  = & \begin{bmatrix}
    0 & 0 & -g_{sx}^{(\mathrm{I}) yz} \\
    0 & 0 & 0 \\
    -g_{sx}^{(\mathrm{I}) zy} & 0 & 0
  \end{bmatrix}, \label{eq:spin_accumulation_1_altermagnets_y} \\
  g_{sz}^{(\mathrm{I})}
  = & \begin{bmatrix}
    0 & g_{sz}^{(\mathrm{I}) xy} & 0 \\
    -g_{sz}^{(\mathrm{I}) xy} & 0 & 0 \\
    0 & 0 & 0
  \end{bmatrix}, \label{eq:spin_accumulation_1_altermagnets_z}
\end{align}\label{eq:spin_accumulation_1_altermagnets}\end{subequations}
for the conventional part and
\begin{subequations}\begin{align}
  g_{sx}^{(\mathrm{II})}
  = & \begin{bmatrix}
    0 & 0 & 0 \\
    0 & 0 & g_{sx}^{(\mathrm{II}) yz} \\
    0 & g_{sx}^{(\mathrm{II}) zy} & 0
  \end{bmatrix}, \label{eq:spin_accumulation_2_altermagnets_x} \\
  g_{sy}^{(\mathrm{II})}
  = & \begin{bmatrix}
    0 & 0 & g_{sx}^{(\mathrm{II}) yz} \\
    0 & 0 & 0 \\
    g_{sx}^{(\mathrm{II}) zy} & 0 & 0
  \end{bmatrix}, \label{eq:spin_accumulation_2_altermagnets_y} \\
  g_{sz}^{(\mathrm{II})}
  = & \begin{bmatrix}
    0 & g_{sz}^{(\mathrm{II}) xy} & 0 \\
    g_{sz}^{(\mathrm{II}) xy} & 0 & 0 \\
    0 & 0 & 0
  \end{bmatrix}, \label{eq:spin_accumulation_2_altermagnets_z}
\end{align}\label{eq:spin_accumulation_2_altermagnets}\end{subequations}
for the magnetic part.
Note that the conventional part is not always antisymmetric with respect to $i$ and $j$,
namely, $g_{sx}^{(\mathrm{I}) zy} \not= -g_{sx}^{(\mathrm{I}) yz}$,
and that the magnetic part is not always symmetric, namely, $g_{sx}^{(\mathrm{II}) zy} \not= g_{sx}^{(\mathrm{II}) yz}$.

We focus on a one-orbital tight-binding model constructed in Refs.~\cite{PhysRevB.110.144412,PhysRevLett.134.096703},
\begin{equation}
  \hat{H}(\bm{k})
  = \epsilon(\bm{k}) + t_{z}(\bm{k}) \tau_{z} + t_{x}(\bm{k}) \tau_{x}
  + \bm{\lambda}(\bm{k}) \cdot \tau_{y} \bm{\sigma} + J \tau_{z} \sigma_{z}, \label{eq:altermagnets_1}
\end{equation}
in which $\bm{\tau}$ and $\bm{\sigma}$ are the Pauli matrices for the sublattice and spin degrees of freedom, respectively,
\begin{subequations}\begin{align}
  \epsilon(\bm{k})
  = & t_{1} (\cos k_{x} a + \cos k_{y} a) + t_{2} \cos k_{z} c \notag \\
  & + t_{3} \cos k_{x} a \cos k_{y} a \notag \\
  & + t_{4} (\cos k_{x} a + \cos k_{y} a) \cos k_{z} c \notag \\
  & + t_{5} \cos k_{x} a \cos k_{y} a \cos k_{z} c - \epsilon_{0}, \label{eq:altermagnets_2a} \\
  t_{z}(\bm{k})
  = & t_{6} \sin k_{x} a \sin k_{y} a
  + t_{7} \sin k_{x} a \sin k_{y} a \cos k_{z} c, \label{eq:altermagnets_2b} \\
  t_{x}(\bm{k})
  = & t_{8} \cos \frac{k_{x} a}{2} \cos \frac{k_{y} a}{2} \cos \frac{k_{z} c}{2}, \label{eq:altermagnets_2c} \\
  \lambda_{x}(\bm{k})
  = & \lambda_{1} \sin \frac{k_{x} a}{2} \cos \frac{k_{y} a}{2} \sin \frac{k_{z} c}{2}, \label{eq:altermagnets_2d} \\
  \lambda_{y}(\bm{k})
  = & -\lambda_{1} \cos \frac{k_{x} a}{2} \sin \frac{k_{y} a}{2} \sin \frac{k_{z} c}{2}, \label{eq:altermagnets_2e} \\
  \lambda_{z}(\bm{k})
  = & \lambda_{2} \cos \frac{k_{x} a}{2} \cos \frac{k_{y} a}{2} \cos \frac{k_{z} c}{2}
  (\cos k_{x} a - \cos k_{y} a), \label{eq:altermagnets_2f}
\end{align}\label{eq:altermagnets_2}\end{subequations}
and $J$ describes the collinear AFM order along the $z$ direction.
$t$'s and $\lambda$'s are the transfer integrals and SOCs, respectively, and $a$ and $c$ are the lattice constants.
Note that $t_{3}, t_{4}, t_{5}$ in Eq.~\eqref{eq:altermagnets_2a} and $t_{7}$ in Eq.~\eqref{eq:altermagnets_2b}
are not taken into account in Ref.~\cite{PhysRevLett.134.096703}.

Below, we compute the intrinsic term $g_{sa}^{(\mathrm{II}) ij}$ in the magnetic part
as well as the conventional part $\gamma_{sa}^{\phantom{sa} ij} = -g_{sa}^{(\mathrm{I}) ij} / \tau$,
in which $\tau = \hbar /\eta$ is the phenomenological relaxation time.
For comparison, we also compute $g_{sz}^{(\mathrm{II}) ij}$ in the absence of the SOCs.
We deal with the AFM order parameter $J$ as a constant but not determined self-consistently.

\subsection{RuO$_{2}$} \label{sub:RuO2}
\begin{figure}
  \centering
  \includegraphics[clip,width=0.48\textwidth]{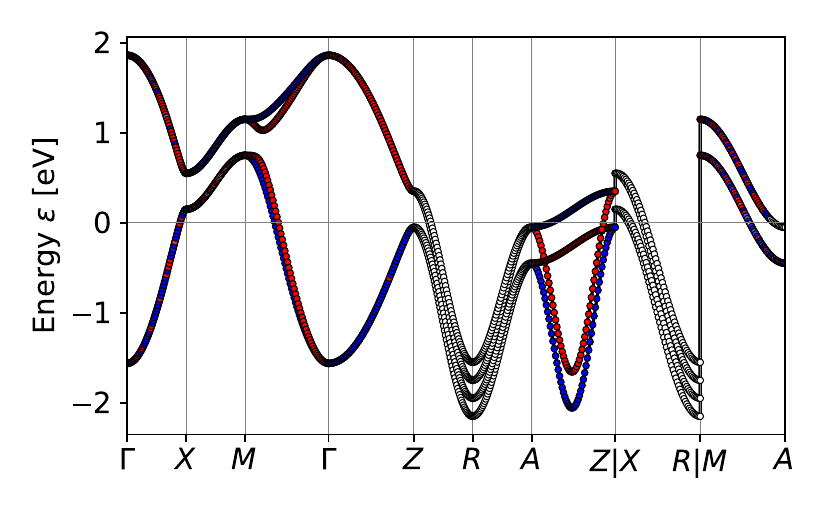}
  \caption{%
    Band structure for RuO$_{2}$.
    Red and blue circles represent spin-up and spin-down dominant bands, respectively.
    Altermagnetic spin splitting is found in $M$-$\Gamma$ and $A$-$Z$ lines.%
  } \label{fig:RuO2_w_soc_w_afm_band}
\end{figure}

\begin{table*}
  \centering
  \caption{%
    Parameters for RuO$_{2}$ and MnF$_{2}$~\cite{PhysRevB.110.144412}.
    $t$'s, $\epsilon_{0}$, $\lambda_{1} = \lambda_{2}$, and $J$ are in the unit of eV, and $a$ and $c$ are in that of \AA.
    For MnF$_{2}$, $\lambda_{1} = \lambda_{2}$ highlighted with $\ast$ are estimated from Ref.~\cite{PhysRevB.102.014422},
    and $J$ highlighted with $\dagger$ is chosen to make the system insulating.%
  } \label{tab:altermagnets}
  \begin{tabular}{cccccccccccccc} \hline \hline
    & $t_{1}$ & $t_{2}$ & $t_{3}$ & $t_{4}$ & $t_{5}$ & $t_{6}$ & $t_{7}$ & $t_{8}$ & $\epsilon_{0}$ & $\lambda_{1} = \lambda_{2}$ & $J$ & $a$ & $c$ \\ \hline
    RuO$_{2}$ & $-0.05$ & $0.7$ & $0.5$ & $-0.15$ & $-0.4$ & $-0.6$ & $0.3$ & $1.7$ & $0.25$ & $0.1$ & $0.2$ & $4.48$ & $3.11$ \\
    MnF$_{2}$ & $0$ & $0.13$ & $0$ & $-0.02$ & $0.015$ & $0$ & $0.03$ & $0.33$ & $-0.01$ & $0.0085^{\ast}$ & $0.3^{\dagger}$ & $4.87$ & $3.31$ \\ \hline \hline
  \end{tabular}
\end{table*}
We use the parameters listed in the upper row of Table~\ref{tab:altermagnets}.
We reproduce the altermagnetic spin splitting in $M$-$\Gamma$ and $A$-$Z$ lines~\cite{PhysRevB.110.144412},
as shown in Fig.~\ref{fig:RuO2_w_soc_w_afm_band}.
The number of $\bm{k}$ points is fixed to $144^{3}$.

\begin{figure}
  \centering
  \includegraphics[clip,width=0.48\textwidth]{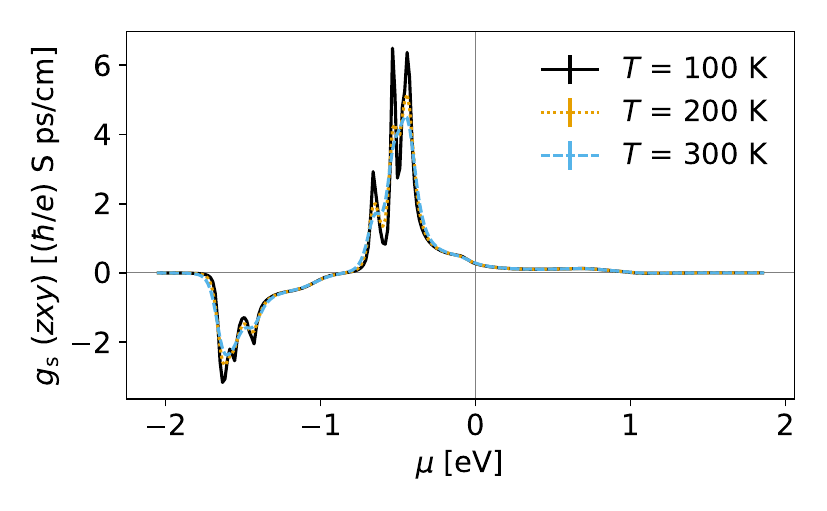}
  \caption{%
    Chemical potential dependence of the intrinsic SAC for RuO$_{2}$ in the absence of the SOCs.
    Only $g_{sz}^{(\mathrm{II}) xy} = g_{sz}^{(\mathrm{II}) yx}$ is allowed.%
  } \label{fig:RuO2_wo_soc_w_afm_mshe}
\end{figure}

\begin{figure*}
  \centering
  \includegraphics[clip,width=0.98\textwidth]{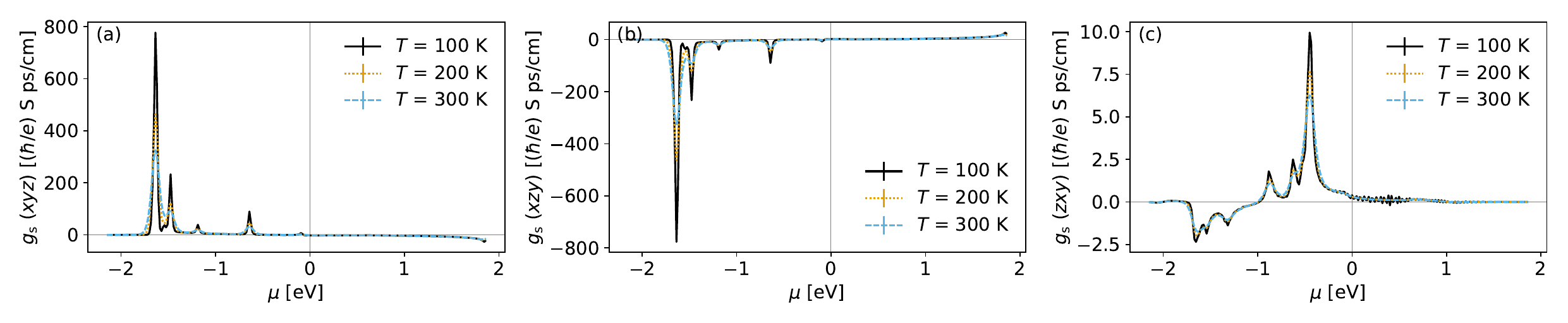}
  \caption{%
    Chemical potential dependence of the intrinsic SAC for RuO$_{2}$.
    (a) $g_{sx}^{(\mathrm{II}) yz} = g_{sy}^{(\mathrm{II}) xz}$,
    (b) $g_{sx}^{(\mathrm{II}) zy} = g_{sy}^{(\mathrm{II}) zx}$,
    and (c) $g_{sz}^{(\mathrm{II}) xy} = g_{sz}^{(\mathrm{II}) yx}$.%
  } \label{fig:RuO2_w_soc_w_afm_mshe}
\end{figure*}

\begin{figure*}
  \centering
  \includegraphics[clip,width=0.98\textwidth]{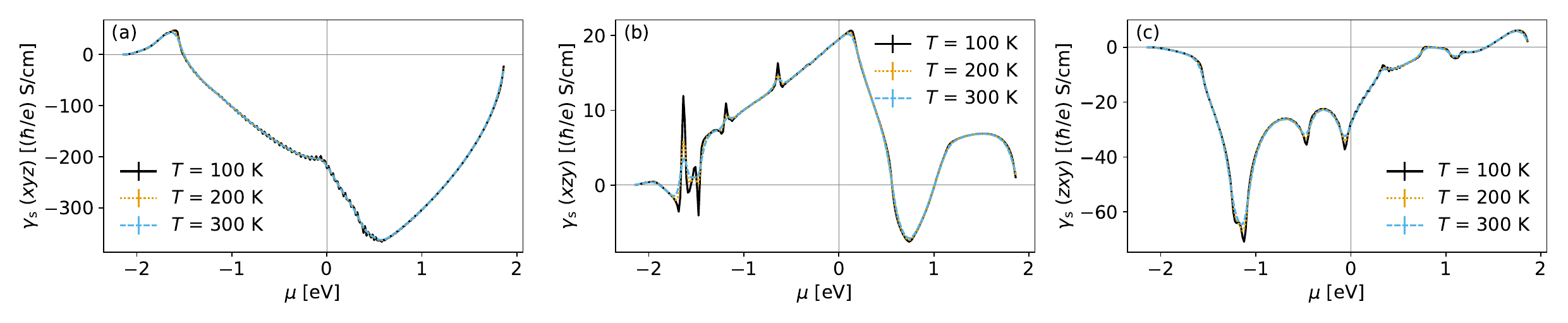}
  \caption{%
    Chemical potential dependence of the conventional SAC for RuO$_{2}$.
    (a) $\gamma_{sx}^{\phantom{sx} yz} = -\gamma_{sy}^{\phantom{sy} xz}$,
    (b) $\gamma_{sx}^{\phantom{sx} zy} = -\gamma_{sy}^{\phantom{sy} zx}$,
    and (c) $\gamma_{sz}^{\phantom{sz} xy} = -\gamma_{sz}^{\phantom{sz} yx}$.%
  } \label{fig:RuO2_w_soc_w_afm_she}
\end{figure*}

The intrinsic term in the absence and presence of the SOCs is shown
in Figs.~\ref{fig:RuO2_wo_soc_w_afm_mshe} and \ref{fig:RuO2_w_soc_w_afm_mshe}, respectively.
Only $g_{sz}^{(\mathrm{II}) xy} = g_{sz}^{(\mathrm{II}) yx}$ is allowed for the former,
while two more components are nonzero for the latter in accordance with Eq.~\eqref{eq:spin_accumulation_2_altermagnets}.
Compared with Fig.~\ref{fig:RuO2_wo_soc_w_afm_mshe},
Fig.~\ref{fig:RuO2_w_soc_w_afm_mshe}(c) is almost unchanged owing to small $\lambda_{1} = \lambda_{2}$.
Notably, $g_{sx}^{(\mathrm{II}) yz} = g_{sy}^{(\mathrm{II}) xz}$ in Fig.~\ref{fig:RuO2_w_soc_w_afm_mshe}(a)
and $g_{sx}^{(\mathrm{II}) zy} = g_{sy}^{(\mathrm{II}) zx}$ in Fig.~\ref{fig:RuO2_w_soc_w_afm_mshe}(b)
are completely antisymmetric with each other beyond the symmetry argument.
Sharp peaks, which come from anticrossing points in the band structure, are fragile against temperature.
Next, we show the conventional part in Fig.~\ref{fig:RuO2_w_soc_w_afm_she}.
Similar to the intrinsic term in Fig.~\ref{fig:RuO2_w_soc_w_afm_mshe}, sharp peaks are fragile against temperature.
Note that $g_{sa}^{(\mathrm{I}) ij} = -\tau \gamma_{sa}^{\phantom{sa} ij}$ may depend on temperature
owing to the relaxation time $\tau$, even though $\gamma_{sa}^{\phantom{sa} ij}$ does not.
We compare these results with the previous experiments in Sec.~\ref{sec:discussion}.

\subsection{MnF$_{2}$} \label{sub:MnF2}
\begin{figure}
  \centering
  \includegraphics[clip,width=0.48\textwidth]{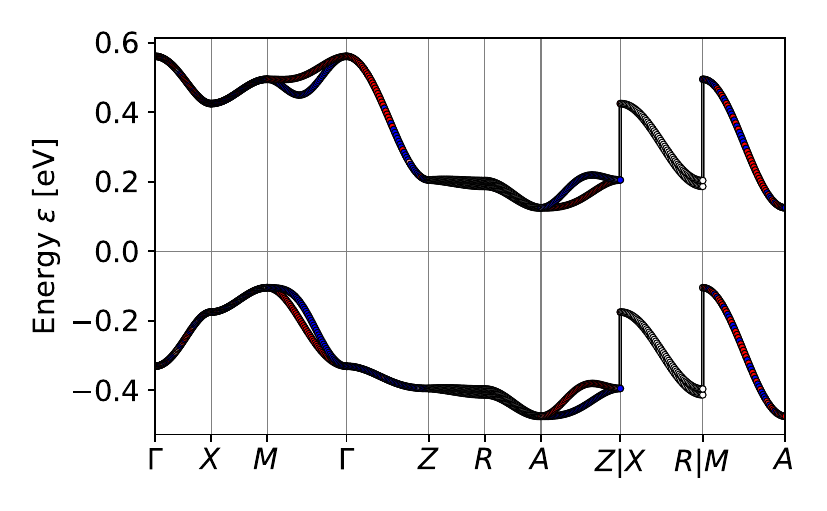}
  \caption{%
    Band structure for MnF$_{2}$.
    Red and blue circles represent spin-up and spin-down dominant bands, respectively.
    Altermagnetic spin splitting is found in $M$-$\Gamma$ and $A$-$Z$ lines.%
  } \label{fig:MnF2_w_soc_w_afm_band}
\end{figure}

We use the parameters listed in the lower row of Table~\ref{tab:altermagnets}.
$\lambda_{1} = \lambda_{2} = 8.5~\mathrm{meV}$ are estimated from the spin splitting at the $R$ point~\cite{PhysRevB.102.014422}.
We choose $J = 0.3~\mathrm{eV}$ to make the system insulating,
but the one-orbital model~\eqref{eq:altermagnets_1} does not reproduce first-principles band calculations~\cite{PhysRevB.110.144412}.
Our aim here is to demonstrate the nonzero SAC in a magnetic insulator.
We show the band structure in Fig.~\ref{fig:MnF2_w_soc_w_afm_band}.
The number of $\bm{k}$ points is fixed to $144^{3}$.

\begin{figure}
  \centering
  \includegraphics[clip,width=0.48\textwidth]{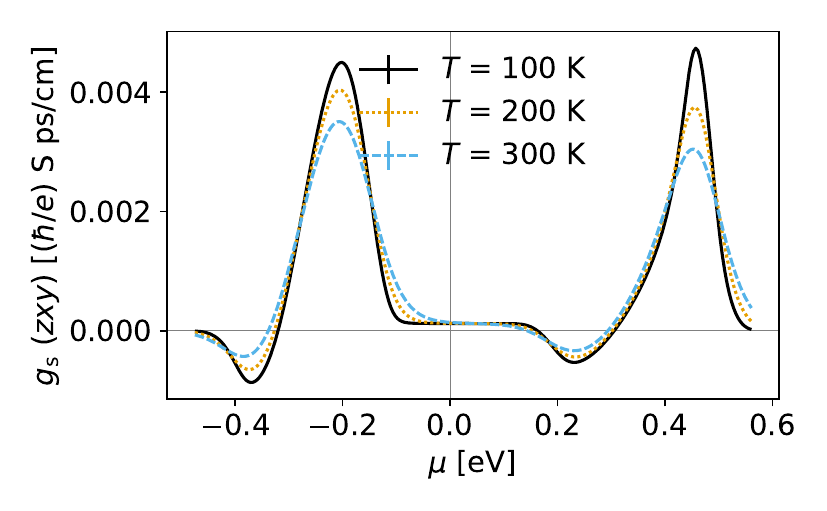}
  \caption{%
    Chemical potential dependence of the intrinsic SAC for MnF$_{2}$ in the absence of the SOCs.
    Only $g_{sz}^{(\mathrm{II}) xy} = g_{sz}^{(\mathrm{II}) yx}$ is allowed.
    We find $g_{sz}^{(\mathrm{II}) xy} = 1.23 \times 10^{-4}~(\hbar/e)~\mathrm{S~ps/cm}$ at the insulating region.%
  } \label{fig:MnF2_wo_soc_w_afm_mshe}
\end{figure}

\begin{figure*}
  \centering
  \includegraphics[clip,width=0.98\textwidth]{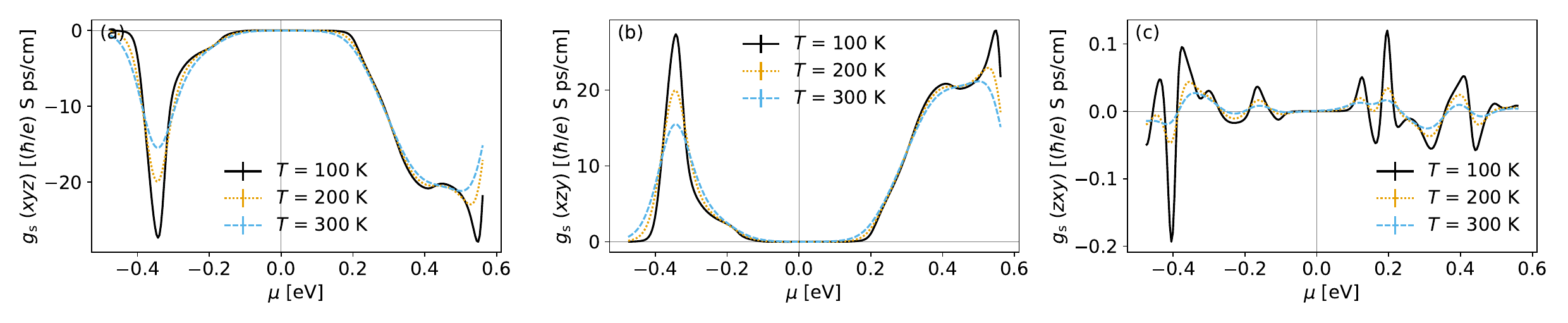}
  \caption{%
    Chemical potential dependence of the intrinsic SAC for MnF$_{2}$.
    (a) $g_{sx}^{(\mathrm{II}) yz} = g_{sy}^{(\mathrm{II}) xz}$,
    (b) $g_{sx}^{(\mathrm{II}) zy} = g_{sy}^{(\mathrm{II}) zx}$,
    and (c) $g_{sz}^{(\mathrm{II}) xy} = g_{sz}^{(\mathrm{II}) yx}$.
    We find $g_{sx}^{(\mathrm{II}) yz} = -1.55 \times 10^{-3}~(\hbar/e)~\mathrm{S~ps/cm}$,
    $g_{sx}^{(\mathrm{II}) zy} = 3.68 \times 10^{-3}~(\hbar/e)~\mathrm{S~ps/cm}$,
    and $g_{sz}^{(\mathrm{II}) xy} = 1.28 \times 10^{-4}~(\hbar/e)~\mathrm{S~ps/cm}$ at the insulating region.%
  } \label{fig:MnF2_w_soc_w_afm_mshe}
\end{figure*}

\begin{figure*}
  \centering
  \includegraphics[clip,width=0.98\textwidth]{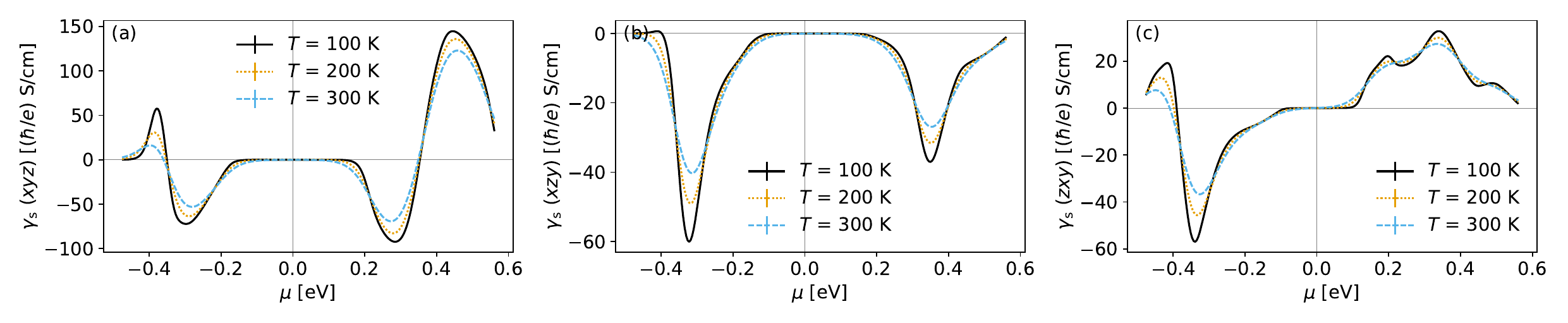}
  \caption{%
   Chemical potential dependence of the conventional SAC for MnF$_{2}$.
    (a) $\gamma_{sx}^{\phantom{sx} yz} = -\gamma_{sy}^{\phantom{sy} xz}$,
    (b) $\gamma_{sx}^{\phantom{sx} zy} = -\gamma_{sy}^{\phantom{sy} zx}$,
    and (c) $\gamma_{sz}^{\phantom{sz} xy} = -\gamma_{sz}^{\phantom{sz} yx}$.%
  } \label{fig:MnF2_w_soc_w_afm_she}
\end{figure*}

The intrinsic term in the absence and presence of the SOCs is shown
in Figs.~\ref{fig:MnF2_wo_soc_w_afm_mshe} and \ref{fig:MnF2_w_soc_w_afm_mshe}, respectively.
For the former, we find small but nonzero $g_{sz}^{(\mathrm{II}) xy} = 1.23 \times 10^{-4}~(\hbar/e)~\mathrm{S~ps/cm}$
at the insulating region.
This magnitude is not affected by the SOCs; $g_{sz}^{(\mathrm{II}) xy} = 1.28 \times 10^{-4}~(\hbar/e)~\mathrm{S~ps/cm}$,
while the other two components turn out to be ten times larger as
$g_{sx}^{(\mathrm{II}) yz} = -1.55 \times 10^{-3}~(\hbar/e)~\mathrm{S~ps/cm}$
and $g_{sx}^{(\mathrm{II}) zy} = 3.68 \times 10^{-3}~(\hbar/e)~\mathrm{S~ps/cm}$.
We emphasize again that the magnetic SH conductivity vanishes in the insulating region.
We also show the conventional part in Fig.~\ref{fig:MnF2_w_soc_w_afm_she}.
This part is extrinsic and vanishes in the insulating region; in other words, the spin response is fully intrinsic.

\section{Discussion and Summary} \label{sec:discussion}
Our theory focuses on the spin accumulation coming from the bulk conventional and magnetic SH effects only.
The whole spin accumulation is sensitive to the surface details, which cannot be described by our bulk theory.
In fact, the Rashba SOC at the surface gives rise to the spin Edelstein and magnetoelectric effects.
On the other hand, when the system is attached to a ferromagnet,
the spin accumulation causes the damping-like spin torque via the exchange coupling at the interface and the Gilbert damping.
These points were discussed in more detail in Ref.~\cite{Shitade2025}.

Here, we mention the previous spin-torque ferromagnetic resonance measurements in RuO$_{2}$~\cite{PhysRevLett.128.197202}.
Following a theoretical proposal~\cite{PhysRevLett.126.127701}, the authors disentangled
the spin-splitter torque $\tau_{\mathrm{SST}} = 130~(\hbar / e)~\mathrm{S/cm}$ owing to the magnetic SH effect
from the spin-orbit torque $\tau_{\mathrm{SOT}} = 70~(\hbar / e)~\mathrm{S/cm}$ owing to the conventional SH effect.
On the other hand, from Figs.~\ref{fig:RuO2_w_soc_w_afm_mshe}(c) and \ref{fig:RuO2_w_soc_w_afm_she}(c), we obtain
$g_{sz}^{(\mathrm{II}) xy} = 0.3~(\hbar / e)~\mathrm{S~ps/cm}$ and $\gamma_{sz}^{\phantom{sz} xy} = -28~(\hbar / e)~\mathrm{S/cm}$
at $\mu = 0$ and $T = 300~\mathrm{K}$.
The former value is converted to $-g_{sz}^{(\mathrm{II}) xy} / \tau = -3~(\hbar / e)~\mathrm{S/cm}$
if $\eta = 6.6~\mathrm{meV}$ ($\tau = 0.1~\mathrm{ps}$) is used~\cite{PhysRevLett.126.127701}.
Thus, our results are reasonable considering that
the one-orbital minimal model does not reproduce first-principles band calculations~\cite{PhysRevB.110.144412}
and the second term of Eq.~\eqref{eq:spin_accumulation_2_copy} is not computed.

The intrinsic SAC in Eq.~\eqref{eq:spin_accumulation_2_copy} is related to the spin magnetic octupole moment.
To see this, we consider a nonuniform chemical potential.
The total orbital or spin magnetization can be decomposed into
\begin{equation}
  M_{a}^{\textrm{tot}}
  = M_{a} - \partial_{x^{j}} M_{a}^{\phantom{a} j} + \partial_{x^{i}} \partial_{x^{j}} M_{a}^{\phantom{a} ij}
  - \dots, \label{eq:multipole}
\end{equation}
in which $M_{a}$, $M_{a}^{\phantom{a} j}$, and $M_{a}^{\phantom{a} ij}$ are the magnetic dipole, quadrupole, and octupole moments,
respectively.
Hence, the uniform part is expressed by
\begin{align}
  M_{a}
  = & M_{a}^{\textrm{tot}} + \partial_{x^{j}} M_{a}^{\phantom{a} j} - \partial_{x^{i}} \partial_{x^{j}} M_{a}^{\phantom{a} ij}
  + \dots \notag \\
  = & M_{a}^{\textrm{tot}} + \partial_{x^{j}} \mu \frac{\partial M_{a}^{\phantom{a} j}}{\partial \mu}
  - \partial_{x^{i}} \partial_{x^{j}} \mu \frac{\partial M_{a}^{\phantom{a} ij}}{\partial \mu}
  - \partial_{x^{i}} \mu \partial_{x^{j}} \mu \frac{\partial^{2} M_{a}^{\phantom{a} ij}}{\partial \mu^{2}}. \label{eq:multipole_2}
\end{align}
Since $q E_{j} - \partial_{x^{j}} \mu = 0$ holds in equilibrium,
we can replace $\partial_{x^{j}} \mu$ with $-q E_{j}$ for nonequilibrium responses.
From the second term of Eq.~\eqref{eq:multipole_2},
we obtain a relation between the magnetoelectric susceptibility and magnetic quadrupole moment as
$\partial M_{a} / \partial E_{j} = -q (\partial M_{a}^{\phantom{a} j} / \partial \mu)$~\cite{PhysRevB.97.134423,PhysRevB.98.020407,PhysRevB.98.060402,PhysRevB.99.024404}.
Similarly, from the third term, we obtain a relation between the accumulation coefficient and magnetic octupole moment as
$\partial M_{a} / \partial (\partial_{x^{i}} E_{j}) = q (\partial M_{a}^{\phantom{a} ij} / \partial \mu)$~\cite{PhysRevB.112.134412,2504.21431}.
These kinds of equations are known as the Streda formula
for the Hall conductivity and orbital magnetization~\cite{0022-3719-15-22-005}.

We can check the above relation by microscopic calculation.
Recently, a generic formula for the spin magnetic octupole moment has been derived
partly by the semiclassical wave packet dynamics~\cite{PhysRevLett.131.106701}
and quantum-mechanically from the thermodynamic definition~\cite{PhysRevB.112.134412,2504.21431}.
We follow the derivation in Appendix~\ref{app:spin_magnetic_octupole}.
In our notation, the formula is expressed by
\begin{widetext}
  \begin{align}
    S_{a}^{\phantom{a} ij}
    = & \frac{1}{2} \sum_{n} \int \frac{d^{d} k}{(2 \pi)^{d}} \left\{
      s_{na}^{\phantom{na} ij} f(\epsilon_{n})
      - \frac{1}{12} s_{na} \partial_{k_{i}} \partial_{k_{j}} \epsilon_{n} f^{\prime}(\epsilon_{n})
      + \sum_{m (\not= n)} \left[
        \frac{i \langle u_{n} | \hbar \hat{v}^{i} | u_{m} \rangle s_{mna}^{\phantom{mna} j} + \cc}{(\epsilon_{n} - \epsilon_{m})^{2}}
        f^{(-1)}(\epsilon_{n})
      \right.
    \right.\notag \\
    & \left. 
      \left.
            - \frac{1}{2} \frac{
          \langle u_{n} | \hbar \hat{v}^{i} | u_{m} \rangle \langle u_{m} | \hat{s}_{a} | u_{n} \rangle + \cc
        }{
          (\epsilon_{n} - \epsilon_{m})^{2}
        } \partial_{k_{j}} \epsilon_{n} f(\epsilon_{n})
      \right]
    \right\} + (i \leftrightarrow j), \label{eq:thermodynamic_octupole_2_copy}
  \end{align}
\end{widetext}
which is by construction symmetric with respect to $i$ and $j$.
Here, we have introduced the intraband spin magnetic octupole moment as
\begin{equation}
  s_{na}^{\phantom{na} ij}
  = \frac{1}{2} \langle D_{k_{i}} u_{n} | \frac{s_{na} + \hat{s}_{a}}{2} | D_{k_{j}} u_{n} \rangle + \cc
  - \frac{1}{6} \partial_{k_{i}} \partial_{k_{j}} s_{na}, \label{eq:spin_magnetic_octupole_copy}
\end{equation}
and
\begin{equation}
  f^{(-1)}(\epsilon)
  = \int_{\infty}^{\epsilon} d z f(z), \label{eq:fermi-1}
\end{equation}
which satisfies $f^{(-1) \prime}(\epsilon) = f(\epsilon)$.
Then, we obtain
\begin{equation}
  \frac{1}{2} (g_{sa}^{(\mathrm{II}) ij} + g_{sa}^{(\mathrm{II}) ji})
  = q \frac{\partial S_{a}^{\phantom{a} ij}}{\partial \mu}, \label{eq:streda}
\end{equation}
in insulators so as to neglect the Fermi-surface terms.
References~\cite{PhysRevB.112.134412,2504.21431} mentioned Eq.~\eqref{eq:streda}
without any microscopic expression~\eqref{eq:spin_accumulation_2_copy}.
The thermodynamic definition of the spin magnetic octupole moment gives the symmetric components only,
and it remains unclear whether Eq.~\eqref{eq:streda} holds for the antisymmetric components as well.

We also comment on the intrinsic spin accumulation for the magnetic spin Nernst effect.
By considering a nonuniform temperature, instead of the chemical potential,
in Eqs.~\eqref{eq:multipole} and \eqref{eq:multipole_2}, we obtain
$\partial M_{a} / \partial (-\partial_{x^{i}} \partial_{x^{j}} T) = \partial M_{a}^{\phantom{a} ij} / \partial T$.
This relation is again valid only for the symmetric components.
Microscopic calculation for the left-hand side, including the antisymmetric components, is a future problem.

To summarize, we have formulated the magnetic SH effect in terms of the SAC,
namely, the linear response of spin to an electric field gradient.
In contrast to the magnetic SH conductivity that vanishes in insulators,
the intrinsic contribution to the SAC is allowed and can be nonzero.
We have derived its generic formula~\eqref{eq:spin_accumulation_2_copy} expressed by Bloch wave functions
and applied it to a minimal model for altermagnets, RuO$_{2}$ and MnF$_{2}$~\cite{PhysRevB.110.144412,PhysRevLett.134.096703}.
In particular, MnF$_{2}$ is insulating,
and only the spin current response for the conventional SH effect and the spin response for the magnetic SH effect are allowed.
We believe that our results stimulate SH experiments on magnetic insulators
to clarify which quantity is primary and experimentally observed.
Finally, we have also shown simple relations of the intrinsic SAC and its thermal counterpart
to the spin magnetic octupole moment~\cite{PhysRevB.112.134412,2504.21431}.

\begin{acknowledgments}
  We thank S.~Hayami, S.~Sumita, and M.~Naka for fruitful discussion on altermagnets and H.~Watanabe on the symmetry argument.
  This work was supported by the Japan Society for the Promotion of Science KAKENHI Grant No.~JP25K07185.
\end{acknowledgments}
\appendix
\begin{widetext}
  \section{Linear Response of Spin to Vector Potential} \label{app:spin_current}
  In this Appendix, we compute the spin--charge-current correlation function,
  \begin{align}
    \chi_{\hat{s}_{a} \hat{J}^{j}}^{\mathrm{R}}(\Omega, \bm{Q})
    = & -\frac{q}{\hbar} \sum_{nm} \int \frac{d^{d} k}{(2 \pi)^{d}}
    \langle u_{n} (\bm{k}_{-}) | \hat{s}_{a} | u_{m}(\bm{k}_{+}) \rangle
    \langle u_{m}(\bm{k}_{+}) | \hbar \hat{v}^{j}(\bm{k}; \bm{Q}) | u_{n}(\bm{k}_{-}) \rangle
    \frac{
      f(\epsilon_{n}(\bm{k}_{-})) - f(\epsilon_{m}(\bm{k}_{+}))
    }{
      \hbar \Omega + \epsilon_{n}(\bm{k}_{-}) - \epsilon_{m}(\bm{k}_{+}) + i \eta
    } \notag \\
    = & \chi_{\hat{s}_{a} \hat{J}^{j}}^{\mathrm{R}}(0, \bm{Q})
    + (i \Omega) \alpha_{\hat{s}_{a} \hat{J}^{j}}^{\mathrm{R}}(\Omega, \bm{Q}), \label{eq:spin_current}
  \end{align}
  which describes the linear response of spin to a vector potential,
  $\langle \Delta \hat{s}_{a} \rangle(\Omega, \bm{Q}) = \chi_{\hat{s}_{a} \hat{J}^{j}}^{\mathrm{R}}(\Omega, \bm{Q}) A_{j}(\Omega, \bm{Q})$.
  Here, $q$ is the electron charge, $d$ is the spatial dimension, $\eta \rightarrow +0$ is the convergence factor,
  $\bm{k}_{\pm} = \bm{k} \pm \bm{Q}/2$, and $f(\epsilon) = [e^{(\epsilon - \mu)/T} + 1]^{-1}$ is the Fermi distribution function.
  $\hat{s}_{a}$ and
  $\hat{v}^{j}(\bm{k}; \bm{Q}) = [\hat{v}^{j}(\bm{k}_{+}) + \hat{v}^{j}(\bm{k}_{-})] / 2$
  are the spin and velocity operators, respectively.
  Since we are interested in the response to an electric field gradient,
  we expand
  \begin{align}
    \alpha_{\hat{s}_{a} \hat{J}^{j}}^{\mathrm{R}}(\Omega, \bm{Q})
    = & \frac{
      \chi_{\hat{s}_{a} \hat{J}^{j}}^{\mathrm{R}}(\Omega, \bm{Q}) - \chi_{\hat{s}_{a} \hat{J}^{j}}^{\mathrm{R}}(0, \bm{Q})
      }{
        i \Omega
      } \notag \\
    = & -i q \sum_{nm} \int \frac{d^{d} k}{(2 \pi)^{d}}
    \langle u_{n} (\bm{k}_{-}) | \hat{s}_{a} | u_{m}(\bm{k}_{+}) \rangle
    \langle u_{m}(\bm{k}_{+}) | \hbar \hat{v}^{j}(\bm{k}; \bm{Q}) | u_{n}(\bm{k}_{-}) \rangle \notag \\
    & \times \frac{
      f(\epsilon_{n}(\bm{k}_{-})) - f(\epsilon_{m}(\bm{k}_{+}))
    }{
      [\hbar \Omega + \epsilon_{n}(\bm{k}_{-}) - \epsilon_{m}(\bm{k}_{+}) + i \eta] [\epsilon_{n}(\bm{k}_{-}) - \epsilon_{m}(\bm{k}_{+})]
    }. \label{eq:spin_current_2}
  \end{align}
  up to the first order with respect to $\bm{Q}$.
  This calculation was already done for the intraband contribution in Ref.~\cite{PhysRevB.105.L201202}.
  Below, we demonstrate more sophisticated derivation both for the intraband and interband contributions.
  
  First, we expand the distribution functions in Eq.~\eqref{eq:spin_current_2}
  with $\epsilon_{n}(\bm{k}_{\pm}) = \epsilon_{n} \pm \delta_{n} + \Delta_{n}$.
  Hereafter, we omit the argument of $\bm{k}$ for simplicity.
  $\delta_{n} = (Q_{j} / 2) \partial_{k_{j}} \epsilon_{n}$ is of the first order with respect to $\bm{Q}$,
  while $\Delta_{n} = (Q_{i} Q_{j} / 8) \partial_{k_{i}} \partial_{k_{j}} \epsilon_{n}$ is of the second order.
  For $n = m$, we obtain
  \begin{align}
    \frac{
      f(\epsilon_{n}(\bm{k}_{-})) - f(\epsilon_{n}(\bm{k}_{+}))
    }{
      [\hbar \Omega + \epsilon_{n}(\bm{k}_{-}) - \epsilon_{n}(\bm{k}_{+}) + i \eta] [\epsilon_{n}(\bm{k}_{-}) - \epsilon_{n}(\bm{k}_{+})]
    }
    = & \frac{
      f(\epsilon_{n} - \delta_{n} + \Delta_{n}) - f(\epsilon_{n} + \delta_{n} + \Delta_{n})
    }{
      (\hbar \Omega + i \eta - 2 \delta_{n}) (-2 \delta_{n})
    } \notag \\
    = & \left[
      \frac{1}{\hbar \Omega + i \eta} + \frac{2 \delta_{n}}{(\hbar \Omega + i \eta)^{2}}
    \right] f^{\prime}(\epsilon_{n}) \notag \\
    = & \frac{f^{\prime}(\epsilon_{n})}{\hbar \Omega + i \eta}
    + Q_{i} \partial_{k_{i}} \epsilon_{n} \frac{f^{\prime}(\epsilon_{n})}{(\hbar \Omega + i \eta)^{2}}, \label{eq:dynamical_intraband}
  \end{align}
  while for $n \not= m$, we obtain
  \begin{align}
    & \frac{
      f(\epsilon_{n}(\bm{k}_{-})) - f(\epsilon_{m}(\bm{k}_{+}))
    }{
      [\hbar \Omega + \epsilon_{n}(\bm{k}_{-}) - \epsilon_{m}(\bm{k}_{+}) + i \eta] [\epsilon_{n}(\bm{k}_{-}) - \epsilon_{m}(\bm{k}_{+})]
    } \notag \\
    = & \frac{
      f(\epsilon_{n} - \delta_{n}) - f(\epsilon_{m} + \delta_{m})
    }{
      (\hbar \Omega + \epsilon_{n} - \epsilon_{m} + i \eta - \delta_{n} - \delta_{m}) (\epsilon_{n} - \epsilon_{m} - \delta_{n} - \delta_{m})
    } \notag \\
    = & \left(
      1 + \frac{\delta_{n} + \delta_{m}}{\hbar \Omega + \epsilon_{n} - \epsilon_{m} + i \eta}
    \right) \left(
      1 + \frac{\delta_{n} + \delta_{m}}{\epsilon_{n} - \epsilon_{m}}
    \right) \frac{
      f(\epsilon_{n}) - f(\epsilon_{m}) - \delta_{n} f^{\prime}(\epsilon_{n}) - \delta_{m} f^{\prime}(\epsilon_{m})
    }{
        (\hbar \Omega + \epsilon_{n} - \epsilon_{m} + i \eta) (\epsilon_{n} - \epsilon_{m})
    } \notag \\
    = & \frac{f(\epsilon_{n}) - f(\epsilon_{m})}{(\hbar \Omega + \epsilon_{n} - \epsilon_{m} + i \eta) (\epsilon_{n} - \epsilon_{m})}
    + \frac{Q_{i}}{2} \left[
      \partial_{k_{i}} (\epsilon_{n} + \epsilon_{m}) \left(
        \frac{1}{\hbar \Omega + \epsilon_{n} - \epsilon_{m} + i \eta} + \frac{1}{\epsilon_{n} - \epsilon_{m}}
      \right) \frac{f(\epsilon_{n}) - f(\epsilon_{m})}{(\hbar \Omega + \epsilon_{n} - \epsilon_{m} + i \eta) (\epsilon_{n} - \epsilon_{m})}
    \right. \notag \\
    & \left.
      - \frac{
        \partial_{k_{i}} \epsilon_{n} f^{\prime}(\epsilon_{n}) + \partial_{k_{i}} \epsilon_{m} f^{\prime}(\epsilon_{m})
      }{
        (\hbar \Omega + \epsilon_{n} - \epsilon_{m} + i \eta) (\epsilon_{n} - \epsilon_{m})
      }
    \right]. \label{eq:dynamical_interband}
  \end{align}
  
  Second, we expand the matrix elements in Eq.~\eqref{eq:spin_current_2} as
  \begin{align}
    & \langle u_{n} (\bm{k}_{-}) | \hat{s}_{a} | u_{m}(\bm{k}_{+}) \rangle
    \langle u_{m}(\bm{k}_{+}) | \hbar \hat{v}^{j}(\bm{k}; \bm{Q}) | u_{n}(\bm{k}_{-}) \rangle \notag \\
    = & \tr [\hat{s}_{a} | u_{m}(\bm{k}_{+}) \rangle \langle u_{m}(\bm{k}_{+}) |
    \hbar \hat{v}^{j}(\bm{k}; \bm{Q}) | u_{n}(\bm{k}_{-}) \rangle \langle u_{n} (\bm{k}_{-}) |] \notag \\
    = & \tr \left\{
      \hat{s}_{a} \left[
        | u_{m} \rangle \langle u_{m} | + \frac{Q_{i}}{2} \partial_{k_{i}} (| u_{m} \rangle \langle u_{m} |)
      \right] \hbar \hat{v}^{j} \left[
        | u_{n} \rangle \langle u_{n} | - \frac{Q_{i}}{2} \partial_{k_{i}} (| u_{n} \rangle \langle u_{n} |)
      \right]
    \right\} \notag \\
    = & \tr (\hat{s}_{a} | u_{m} \rangle \langle u_{m} | \hbar \hat{v}^{j} | u_{n} \rangle \langle u_{n} |)
    + \frac{Q_{i}}{2} \tr [
      \hat{s}_{a} \partial_{k_{i}} (| u_{m} \rangle \langle u_{m} |) \hbar \hat{v}^{j} | u_{n} \rangle \langle u_{n} |
      - \hat{s}_{a} | u_{m} \rangle \langle u_{m} | \hbar \hat{v}^{j} \partial_{k_{i}} (| u_{n} \rangle \langle u_{n} |)
    ]. \label{eq:spin_current_3}
  \end{align}
  Here, we introduce the covariant derivative as
  $| D_{k_{i}} u_{n} \rangle  = | \partial_{k_{i}} u_{n} \rangle + i | u_{n} \rangle a_{n}^{i}$,
  where $a_{n}^{i} = i \langle u_{n} | \partial_{k_{i}} u_{n} \rangle$ is the intraband Berry connection.
  By construction, $\langle u_{n} | D_{k_{i}} u_{n} \rangle = 0$.
  Since $| u_{n} \rangle \langle u_{n} |$ is gauge-covariant, its derivative is as well.
  Hence, we find
  $\partial_{k_{i}} (| u_{n} \rangle \langle u_{n} |) = | D_{k_{i}} u_{n} \rangle \langle u_{n} | + | u_{n} \rangle \langle D_{k_{i}} u_{n} |$.
  Using this identity, we obtain
  \begin{align}
    \text{Eq.~\eqref{eq:spin_current_3}}
    = & \tr (\hat{s}_{a} | u_{m} \rangle \langle u_{m} | \hbar \hat{v}^{j} | u_{n} \rangle \langle u_{n} |)
    + \frac{Q_{i}}{2} \tr [
      \hat{s}_{a} (| D_{k_{i}} u_{m} \rangle \langle u_{m} | + | u_{m} \rangle \langle D_{k_{i}} u_{m} |)
      \hbar \hat{v}^{j} | u_{n} \rangle \langle u_{n} | \notag \\
      & - \hat{s}_{a} | u_{m} \rangle \langle u_{m} |
      \hbar \hat{v}^{j} (| D_{k_{i}} u_{n} \rangle \langle u_{n} | + | u_{n} \rangle \langle D_{k_{i}} u_{n} |)
    ] \notag \\
    = & \langle u_{n} | \hat{s}_{a} | u_{m} \rangle \langle u_{m} | \hbar \hat{v}^{j} | u_{n} \rangle
    + \frac{Q_{i}}{2} [
      (\langle u_{n} | \hat{s}_{a} | D_{k_{i}} u_{m} \rangle - \langle D_{k_{i}} u_{n} | \hat{s}_{a} | u_{m} \rangle)
      \langle u_{m} | \hbar \hat{v}^{j} | u_{n} \rangle \notag \\
      & - \langle u_{n} | \hat{s}_{a} | u_{m} \rangle
      (\langle u_{m} | \hbar \hat{v}^{j} | D_{k_{i}} u_{n} \rangle - \langle D_{k_{i}} u_{m} | \hbar \hat{v}^{j} | u_{n} \rangle)
    ]. \label{eq:spin_current_4}
  \end{align}

  To calculate Eq.~\eqref{eq:spin_current_4} further, we consider $\tilde{H} = \hat{H} + \hb{s} \cdot \bm{B}$
  and take the $\bm{B} \rightarrow 0$ limit at the end of the calculation.
  By differenciating
  $\langle \tilde{u}_{m} | \tilde{H} = \langle \tilde{u}_{m} | \tilde{\epsilon}_{m}$
  with respect to $k_{j}$ and $B^{a}$, we obtain
  $\langle \tilde{u}_{m} | \hbar \hat{v}^{j}
  = \langle \tilde{u}_{m} | \partial_{k_{j}} \tilde{\epsilon}_{m} + \langle D_{k_{j}} \tilde{u}_{m} | (\tilde{\epsilon}_{m} - \tilde{H})$
  and
  $\langle \tilde{u}_{m} | \hat{s}_{a}
  = \langle \tilde{u}_{m} | \partial_{B^{a}} \tilde{\epsilon}_{m} + \langle D_{B^{a}} \tilde{u}_{m} | (\tilde{\epsilon}_{m} - \tilde{H})$,
  respectively.
  Then, we obtain the famous identities:
  \begin{subequations}\begin{align}
    \langle \tilde{u}_{m} | \hbar \hat{v}^{j} | \tilde{u}_{n} \rangle
    = & \partial_{k_{j}} \tilde{\epsilon}_{n} \delta_{mn}
    + (\tilde{\epsilon}_{n} - \tilde{\epsilon}_{m}) \langle \tilde{u}_{m} | D_{k_{j}} \tilde{u}_{n} \rangle, \label{eq:matrix_element_1} \\
    \langle \tilde{u}_{m} | \hbar \hat{v}^{j} | D_{B^{a}} \tilde{u}_{n} \rangle
    = & \partial_{k_{j}} \tilde{\epsilon}_{m} \langle \tilde{u}_{m} | D_{B^{a}} \tilde{u}_{n} \rangle
    + \langle D_{k_{j}} \tilde{u}_{m} | (\tilde{\epsilon}_{m} - \tilde{H}) | D_{B^{a}} \tilde{u}_{n} \rangle, \label{eq:matrix_element_2} \\
    \langle \tilde{u}_{m} | \hbar \hat{v}^{j} | D_{k_{i}} \tilde{u}_{n} \rangle
    = & \partial_{k_{j}} \tilde{\epsilon}_{m} \langle \tilde{u}_{m} | D_{k_{i}} \tilde{u}_{n} \rangle
    + \langle D_{k_{j}} \tilde{u}_{m} | (\tilde{\epsilon}_{m} - \tilde{H}) | D_{k_{i}} \tilde{u}_{n} \rangle, \label{eq:matrix_element_3} \\
    \langle \tilde{u}_{m} | \hat{s}_{a} | \tilde{u}_{n} \rangle
    = & \partial_{B^{a}} \tilde{\epsilon}_{n} \delta_{mn}
    + (\tilde{\epsilon}_{n} - \tilde{\epsilon}_{m}) \langle \tilde{u}_{m} | D_{B^{a}} \tilde{u}_{n} \rangle, \label{eq:matrix_element_4} \\
    \langle \tilde{u}_{m} | \hat{s}_{a} | D_{k_{j}} \tilde{u}_{n} \rangle
    = & \partial_{B^{a}} \tilde{\epsilon}_{m} \langle \tilde{u}_{m} | D_{k_{j}} \tilde{u}_{n} \rangle
    + \langle D_{B^{a}} \tilde{u}_{m} | (\tilde{\epsilon}_{m} - \tilde{H}) | D_{k_{j}} \tilde{u}_{n} \rangle. \label{eq:matrix_element_5}
  \end{align}\label{eq:matrix_element}\end{subequations}
  Here, we define the interband spin magnetic quadrupole and
  orbital magnetic moments~\cite{PhysRevB.98.060402,PhysRevB.103.045401,PhysRevB.107.094106} as
  \begin{subequations}\begin{align}
    \tilde{s}_{mna}^{\phantom{mna} j}
    = & \frac{i}{2} [
      \langle \tilde{u}_{m} | (\partial_{B^{a}} \tilde{\epsilon}_{n} + \hat{s}_{a}) | D_{k_{j}} \tilde{u}_{n} \rangle
      - \langle \tilde{u}_{m} | (\partial_{k_{j}} \tilde{\epsilon}_{n} + \hbar \hat{v}^{j}) | D_{B^{a}} \tilde{u}_{n} \rangle
    ] \notag \\
    = & \frac{i}{2} [
      \partial_{B^{a}} (\tilde{\epsilon}_{n} + \tilde{\epsilon}_{m}) \langle \tilde{u}_{m} | D_{k_{j}} \tilde{u}_{n} \rangle
      + \langle D_{B^{a}} \tilde{u}_{m} | (\tilde{\epsilon}_{m} - \tilde{H}) | D_{k_{j}} \tilde{u}_{n} \rangle \notag \\
      & - \partial_{k_{j}} (\tilde{\epsilon}_{n} + \tilde{\epsilon}_{m}) \langle \tilde{u}_{m} | D_{B^{a}} \tilde{u}_{n} \rangle
      - \langle D_{k_{j}} \tilde{u}_{m} | (\tilde{\epsilon}_{m} - \tilde{H}) | D_{B^{a}} \tilde{u}_{n} \rangle
    ] \notag \\
    = & \frac{i}{2} \sum_{l (\not= n)} \frac{
      (\tilde{s}_{na} \delta_{ml} + \langle \tilde{u}_{m} | \hat{s}_{a} | \tilde{u}_{l} \rangle)
      \langle \tilde{u}_{l} | \hbar \hat{v}^{j} | \tilde{u}_{n} \rangle
      - (\partial_{k_{j}} \tilde{\epsilon}_{n} \delta_{ml} + \langle \tilde{u}_{m} | \hbar \hat{v}^{j} | \tilde{u}_{l} \rangle)
      \langle \tilde{u}_{l} | \hat{s}_{a} | \tilde{u}_{n} \rangle
    }{
      \tilde{\epsilon}_{n} - \tilde{\epsilon}_{l}
    }, \label{eq:spin_magnetic_quadrupole} \\
    \epsilon^{ija} \tilde{m}_{mna}
    = & -\frac{i}{2} [
      \langle \tilde{u}_{m} | (\partial_{k_{i}} \tilde{\epsilon}_{n} + \hbar \hat{v}^{i}) | D_{k_{j}} \tilde{u}_{n} \rangle
      - (i \leftrightarrow j)
    ] \notag \\
    = & -\frac{i}{2} [
      \partial_{k_{i}} (\tilde{\epsilon}_{n} + \tilde{\epsilon}_{m}) \langle \tilde{u}_{m} | D_{k_{j}} \tilde{u}_{n} \rangle
      + \langle D_{k_{i}} \tilde{u}_{m} | (\tilde{\epsilon}_{m} - \tilde{H}) | D_{k_{j}} \tilde{u}_{n} \rangle
      - (i \leftrightarrow j)
    ] \notag \\
    = & -\frac{i}{2} \sum_{l (\not= n)} \frac{
      (\partial_{k_{i}} \tilde{\epsilon}_{n} \delta_{ml} + \langle \tilde{u}_{m} | \hbar \hat{v}^{i} | \tilde{u}_{l} \rangle)
      \langle \tilde{u}_{l} | \hbar \hat{v}^{j} | \tilde{u}_{n} \rangle
      - (i \leftrightarrow j)
    }{
      \tilde{\epsilon}_{n} - \tilde{\epsilon}_{l}
    }, \label{eq:orbital_magnetic}
  \end{align}\label{eq:magnetic_multipole}\end{subequations}
  using Eq.~\eqref{eq:matrix_element} and $\tilde{s}_{na} = \langle \tilde{u}_{n} | \hat{s}_{a} | \tilde{u}_{n} \rangle$.
  Note that the sign of Eq.~\eqref{eq:orbital_magnetic} is opposite to
  that in Refs.~\cite{PhysRevB.98.060402,PhysRevB.103.045401,PhysRevB.107.094106}.
  The third line of each equation,
  obtained from the first line by inserting $\sum_{l} | \tilde{u}_{l} \rangle \langle \tilde{u}_{l} | = 1$,
  is useful for taking the $\bm{B} \rightarrow 0$ limit.
  These quantities are non-Hermitian as
  \begin{subequations}\begin{align}
    \tilde{s}_{nma}^{\phantom{nma} j \ast}
    = & \frac{i}{2} [
      \partial_{B^{a}} (\tilde{\epsilon}_{n} + \tilde{\epsilon}_{m}) \langle \tilde{u}_{m} | D_{k_{j}} \tilde{u}_{n} \rangle
      + \langle D_{B^{a}} \tilde{u}_{m} | (\tilde{\epsilon}_{n} - \tilde{H}) | D_{k_{j}} \tilde{u}_{n} \rangle
      - \partial_{k_{j}} (\tilde{\epsilon}_{n} + \tilde{\epsilon}_{m}) \langle \tilde{u}_{m} | D_{B^{a}} \tilde{u}_{n} \rangle
      - \langle D_{k_{j}} \tilde{u}_{m} | (\tilde{\epsilon}_{n} - \tilde{H}) | D_{B^{a}} \tilde{u}_{n} \rangle
    ] \notag \\
    = & \tilde{s}_{mna}^{\phantom{mna} j}
    + \frac{i}{2} (\tilde{\epsilon}_{n} - \tilde{\epsilon}_{m})
    (\langle D_{B^{a}} \tilde{u}_{m} | D_{k_{j}} \tilde{u}_{n} \rangle - \langle D_{k_{j}} \tilde{u}_{m} | D_{B^{a}} \tilde{u}_{n} \rangle),
    \label{eq:spin_magnetic_quadrupole_nonhermitian} \\
    \epsilon^{ija} \tilde{m}_{nma}^{\ast}
    = & -\frac{i}{2} [
      \partial_{k_{i}} (\tilde{\epsilon}_{n} + \tilde{\epsilon}_{m}) \langle \tilde{u}_{m} | D_{k_{j}} \tilde{u}_{n} \rangle
      + \langle D_{k_{i}} \tilde{u}_{m} | (\tilde{\epsilon}_{n} - \tilde{H}) | D_{k_{j}} \tilde{u}_{n} \rangle
      - (i \leftrightarrow j)
    ] \notag \\
    = & \epsilon^{ija} \tilde{m}_{mna}
    - \frac{i}{2} (\tilde{\epsilon}_{n} - \tilde{\epsilon}_{m})
    [\langle D_{k_{i}} \tilde{u}_{m} | D_{k_{j}} \tilde{u}_{n} \rangle - (i \leftrightarrow j)]. \label{eq:orbital_magnetic_nonhermitian}
  \end{align}\label{eq:nonhermitian}\end{subequations}
  Nonetheless, their diagonal elements are real
  and coincide with the intraband spin magnetic quadrupole~\cite{PhysRevB.99.024404,PhysRevB.105.L201202}
  and orbital magnetic moments, respectively, as
  \begin{equation}
    \tilde{s}_{na}^{\phantom{na} j}
    = \frac{i}{2} [
      \langle D_{B^{a}} \tilde{u}_{n} | (\tilde{\epsilon}_{n} - \tilde{H}) | D_{k_{j}} \tilde{u}_{n} \rangle
      - \langle D_{k_{j}} \tilde{u}_{n} | (\tilde{\epsilon}_{n} - \tilde{H}) | D_{B^{a}} \tilde{u}_{n} \rangle
    ]
    = \frac{i}{2} (
      \langle \tilde{u}_{n} | \hat{s}_{a} | D_{k_{j}} \tilde{u}_{n} \rangle
      - \langle D_{k_{j}} \tilde{u}_{n} | \hat{s}_{a} | \tilde{u}_{n} \rangle
    ). \label{eq:spin_magnetic_quadrupole_diagonal}
  \end{equation}
  Then we obtain
  \begin{align}
    \text{Eq.~\eqref{eq:spin_current_4}}
    \xleftarrow{\bm{B} \rightarrow 0} & [
      \partial_{B^{a}} \tilde{\epsilon}_{n} \delta_{mn}
      + (\tilde{\epsilon}_{n} - \tilde{\epsilon}_{m}) \langle D_{B^{a}} \tilde{u}_{n} | \tilde{u}_{m} \rangle
      ] [
        \partial_{k_{j}} \tilde{\epsilon}_{n} \delta_{mn}
        + (\tilde{\epsilon}_{n} - \tilde{\epsilon}_{m}) \langle \tilde{u}_{m} | D_{k_{j}} \tilde{u}_{n} \rangle
      ] \notag \\
      & + \frac{Q_{i}}{2} \{
        [
          -\partial_{B^{a}} (\tilde{\epsilon}_{n} + \tilde{\epsilon}_{m}) \langle D_{k_{i}} \tilde{u}_{n} | \tilde{u}_{m} \rangle
          + \langle D_{B^{a}} \tilde{u}_{n} | (\tilde{\epsilon}_{n} - \tilde{H}) | D_{k_{i}} \tilde{u}_{m} \rangle
          - \langle D_{k_{i}} \tilde{u}_{n} | (\tilde{\epsilon}_{m} - \tilde{H}) | D_{B^{a}} \tilde{u}_{m} \rangle
        ] \notag \\
        & \times [
          \partial_{k_{j}} \tilde{\epsilon}_{n} \delta_{mn}
          + (\tilde{\epsilon}_{n} - \tilde{\epsilon}_{m}) \langle \tilde{u}_{m} | D_{k_{j}} \tilde{u}_{n} \rangle
        ]
        - [
          \partial_{B^{a}} \tilde{\epsilon}_{n} \delta_{mn}
          + (\tilde{\epsilon}_{n} - \tilde{\epsilon}_{m}) \langle D_{B^{a}} \tilde{u}_{n} | \tilde{u}_{m} \rangle
        ] \notag \\
        & \times [
          \partial_{k_{j}} (\tilde{\epsilon}_{n} + \tilde{\epsilon}_{m}) \langle \tilde{u}_{m} | D_{k_{i}} \tilde{u}_{n} \rangle
          + \langle D_{k_{j}} \tilde{u}_{m} | (\tilde{\epsilon}_{m} - \tilde{H}) | D_{k_{i}} \tilde{u}_{n} \rangle
          - \langle D_{k_{i}} \tilde{u}_{m} | (\tilde{\epsilon}_{n} - \tilde{H}) | D_{k_{j}} \tilde{u}_{n} \rangle
        ]
      \} \notag \\
      = & \partial_{B^{a}} \tilde{\epsilon}_{n} \partial_{k_{j}} \tilde{\epsilon}_{n} \delta_{mn}
      + (\tilde{\epsilon}_{n} - \tilde{\epsilon}_{m})^{2}
      \langle D_{B^{a}} \tilde{u}_{n} | \tilde{u}_{m} \rangle \langle \tilde{u}_{m} | D_{k_{j}} \tilde{u}_{n} \rangle
      + \frac{Q_{i}}{2} \left(
        - 2 i (
          \tilde{s}_{na}^{\phantom{na} i} \partial_{k_{j}} \tilde{\epsilon}_{n}
          - \partial_{B^{a}} \tilde{\epsilon}_{n} \epsilon^{ijb} \tilde{m}_{nb}
        ) \delta_{mn}
      \right. \notag \\
        & + (\tilde{\epsilon}_{n} - \tilde{\epsilon}_{m}) \{
          [
            -2 i \tilde{s}_{nma}^{\phantom{nma} i}
            - \partial_{k_{i}} (\tilde{\epsilon}_{n} + \tilde{\epsilon}_{m}) \langle D_{B^{a}} \tilde{u}_{n} | \tilde{u}_{m} \rangle
            + (\tilde{\epsilon}_{n} - \tilde{\epsilon}_{m}) \langle D_{k_{i}} \tilde{u}_{n} | D_{B^{a}} \tilde{u}_{m} \rangle
          ] \langle \tilde{u}_{m} | D_{k_{j}} \tilde{u}_{n} \rangle \notag \\
      & \left.
          - \langle D_{B^{a}} \tilde{u}_{n} | \tilde{u}_{m} \rangle [
            - 2 i \epsilon^{ijb} \tilde{m}_{mnb}
            + \partial_{k_{i}} (\tilde{\epsilon}_{n} + \tilde{\epsilon}_{m}) \langle \tilde{u}_{m} | D_{k_{j}} \tilde{u}_{n} \rangle
            - (\tilde{\epsilon}_{n} - \tilde{\epsilon}_{m}) \langle D_{k_{i}} \tilde{u}_{m} | D_{k_{j}} \tilde{u}_{n} \rangle
          ]
        \}
      \right). \label{eq:spin_current_5}
  \end{align}
  
  Finally, we combine Eq.~\eqref{eq:spin_current_5} with Eqs.~\eqref{eq:dynamical_intraband} and \eqref{eq:dynamical_interband}.
  We focus on $\Omega \rightarrow 0$.
  The zeroth-order coefficients with respect to $\bm{Q}$ characterize
  $\langle \hat{s}_{a} \rangle(0, \bm{Q}) = \lim_{\Omega \rightarrow 0} \tilde{\alpha}_{sa}^{\phantom{sa} j} (i \Omega) A_{j}(\Omega, \bm{Q})$.
  There are two contributions from $n = m$ and $n \not= m$:
  \begin{subequations}\begin{align}
    \tilde{\alpha}_{sa}^{(\mathrm{I}) j}
    = & \frac{q}{\eta} \sum_{n} \int \frac{d^{d} k}{(2 \pi)^{d}}
    \partial_{B^{a}} \tilde{\epsilon}_{n} \partial_{k_{j}} \tilde{\epsilon}_{n}
    [-f^{\prime}(\tilde{\epsilon}_{n})], \label{eq:magnetoelectric_1} \\
    \tilde{\alpha}_{sa}^{(\mathrm{II}) j}
    = & -q \sum_{n \not= m} \int \frac{d^{d} k}{(2 \pi)^{d}}
    i \langle D_{B^{a}} \tilde{u}_{n} | \tilde{u}_{m} \rangle \langle \tilde{u}_{m} | D_{k_{j}} \tilde{u}_{n} \rangle
    [f(\tilde{\epsilon}_{n}) - f(\tilde{\epsilon}_{m})] \notag \\
    = & -q \sum_{n \not= m} \int \frac{d^{d} k}{(2 \pi)^{d}}
    (i \langle D_{B^{a}} \tilde{u}_{n} | \tilde{u}_{m} \rangle \langle \tilde{u}_{m} | D_{k_{j}} \tilde{u}_{n} \rangle + \cc)
    f(\tilde{\epsilon}_{n}). \label{eq:magnetoelectric_2}
  \end{align}\label{eq:magnetoelectric}\end{subequations}
  The intraband contribution~\eqref{eq:magnetoelectric_1} is identified as the spin Edelstein coefficient,
  whereas the interband one~\eqref{eq:magnetoelectric_2} as the spin magnetoelectric susceptibility.
  It is easy to take the $\bm{B} \rightarrow 0$ limit using Eq.~\eqref{eq:matrix_element_4}.

  The first-order coefficients characterize
  $\langle \hat{s}_{a} \rangle(0, \bm{Q})
  = \lim_{\Omega \rightarrow 0} \tilde{g}_{sa}^{\phantom{sa} ij} (i Q_{i}) (i \Omega) A_{j}(\Omega, \bm{Q})$.
  There are two contributions from $n = m$ and $n \not= m$:
  \begin{subequations}\begin{align}
    \text{intra.}
    = & -\frac{q}{\eta} \sum_{n} \int \frac{d^{d} k}{(2 \pi)^{d}}
    (\tilde{s}_{na}^{\phantom{na} i} \partial_{k_{j}} \tilde{\epsilon}_{n} - \partial_{B^{a}} \tilde{\epsilon}_{n} \epsilon^{ijb} \tilde{m}_{nb})
    [-f^{\prime}(\tilde{\epsilon}_{n})]
    - \frac{q}{\eta^{2}} \sum_{n} \int \frac{d^{d} k}{(2 \pi)^{d}}
    \partial_{B^{a}} \tilde{\epsilon}_{n} \partial_{k_{i}} \tilde{\epsilon}_{n} \partial_{k_{j}} \tilde{\epsilon}_{n}
    [-f^{\prime}(\tilde{\epsilon}_{n})] \notag \\
    = & -\frac{q}{\eta} \sum_{n} \int \frac{d^{d} k}{(2 \pi)^{d}}
    (\tilde{s}_{na}^{\phantom{na} i} \partial_{k_{j}} \tilde{\epsilon}_{n} - \tilde{s}_{na} \epsilon^{ijb} \tilde{m}_{nb})
    [-f^{\prime}(\tilde{\epsilon}_{n})]
    - \frac{q}{\eta^{2}} \sum_{n} \int \frac{d^{d} k}{(2 \pi)^{d}}
    \tilde{s}_{na} \partial_{k_{i}} \tilde{\epsilon}_{n} \partial_{k_{j}} \tilde{\epsilon}_{n}
    [-f^{\prime}(\tilde{\epsilon}_{n})], \label{eq:spin_accumulation_1} \\
    \text{inter.}
    = & -q \sum_{n \not= m} \int \frac{d^{d} k}{(2 \pi)^{d}} \left\{
      \frac{
        -i \tilde{s}_{nma}^{\phantom{nma} i} \langle \tilde{u}_{m} | D_{k_{j}} \tilde{u}_{n} \rangle
        + i \langle D_{B^{a}} \tilde{u}_{n} | \tilde{u}_{m} \rangle \epsilon^{ijb} \tilde{m}_{mnb}
      }{
        \tilde{\epsilon}_{n} - \tilde{\epsilon}_{m}
      } [f(\tilde{\epsilon}_{n}) - f(\tilde{\epsilon}_{m})]
    \right. \notag \\
    & - \frac{1}{2} \langle D_{B^{a}} \tilde{u}_{n} | \tilde{u}_{m} \rangle \langle \tilde{u}_{m} | D_{k_{j}} \tilde{u}_{n} \rangle
    [
      \partial_{k_{i}} \tilde{\epsilon}_{n} f^{\prime}(\tilde{\epsilon}_{n})
      + \partial_{k_{i}} \tilde{\epsilon}_{m} f^{\prime}(\tilde{\epsilon}_{m})
    ] \notag \\
    & \left.
      + \frac{1}{2} (
        \langle D_{B^{a}} \tilde{u}_{n} | \tilde{u}_{m} \rangle \langle D_{k_{i}} \tilde{u}_{m} | D_{k_{j}}\tilde{u}_{n} \rangle
        + \langle D_{k_{i}} \tilde{u}_{n} | D_{B^{a}} \tilde{u}_{m} \rangle \langle \tilde{u}_{m} | D_{k_{j}} \tilde{u}_{n} \rangle
      ) [f(\tilde{\epsilon}_{n}) - f(\tilde{\epsilon}_{m})]
    \right\} \notag \\
    = & -q \sum_{n \not= m} \int \frac{d^{d} k}{(2 \pi)^{d}} \left[
      \frac{
        i \langle D_{k_{j}} \tilde{u}_{n} | \tilde{u}_{m} \rangle \tilde{s}_{mna}^{\phantom{mna} i}
        + i \langle D_{B^{a}} \tilde{u}_{n} | \tilde{u}_{m} \rangle \epsilon^{ijb} \tilde{m}_{mnb} + \cc
      }{
        \tilde{\epsilon}_{n} - \tilde{\epsilon}_{m}
      } f(\tilde{\epsilon}_{n})
    \right. \notag \\
    & - \frac{1}{2} (\langle D_{B^{a}} \tilde{u}_{n} | \tilde{u}_{m} \rangle \langle \tilde{u}_{m} | D_{k_{j}} \tilde{u}_{n} \rangle + \cc)
    \partial_{k_{i}} \tilde{\epsilon}_{n} f^{\prime}(\tilde{\epsilon}_{n}) \notag \\
    & \left.
      + \frac{1}{2} (
        \langle D_{B^{a}} \tilde{u}_{n} | \tilde{u}_{m} \rangle \langle D_{k_{i}} \tilde{u}_{m} | D_{k_{j}} \tilde{u}_{n} \rangle
        + \langle D_{B^{a}} \tilde{u}_{n} |  D_{k_{i}} \tilde{u}_{m} \rangle \langle \tilde{u}_{m} | D_{k_{j}} \tilde{u}_{n} \rangle + \cc
      ) f(\tilde{\epsilon}_{n})
    \right] \notag \\
    = & -q \sum_{n \not= m} \int \frac{d^{d} k}{(2 \pi)^{d}} \left[
      \frac{
        i \langle \tilde{u}_{n} | \hbar \hat{v}^{j} | \tilde{u}_{m} \rangle \tilde{s}_{mna}^{\phantom{mna} i}
        + i \langle \tilde{u}_{n} | \hat{s}_{a} | \tilde{u}_{m} \rangle \epsilon^{ijb} \tilde{m}_{mnb} + \cc
      }{
        (\tilde{\epsilon}_{n} - \tilde{\epsilon}_{m})^{2}
      } f(\tilde{\epsilon}_{n})
    \right. \notag \\
    & \left.
      - \frac{1}{2} \frac{
        \langle \tilde{u}_{n} | \hat{s}_{a} | \tilde{u}_{m} \rangle \langle \tilde{u}_{m} | \hbar \hat{v}^{j} | \tilde{u}_{n} \rangle + \cc
      }{
        (\tilde{\epsilon}_{n} - \tilde{\epsilon}_{m})^{2}
      } \partial_{k_{i}} \tilde{\epsilon}_{n} f^{\prime}(\tilde{\epsilon}_{n})
    \right]. \label{eq:spin_accumulation_2}
  \end{align}\label{eq:spin_accumulation}\end{subequations}
  The final expression of each equation is useful for taking the $\bm{B} \rightarrow 0$ limit.
  The intraband contribution~\eqref{eq:spin_accumulation_1} coincides with the previous result in Ref.~\cite{PhysRevB.105.L201202}.
  In Eq.~\eqref{eq:spin_accumulation_2},
  we have used Eq.~\eqref{eq:nonhermitian} from the first line to the second one.
  The third term in the second line vanishes as
  \begin{equation}
    \sum_{m (\not= n)} (
      \langle D_{B^{a}} \tilde{u}_{n} | \tilde{u}_{m} \rangle \langle D_{k_{i}} \tilde{u}_{m} | D_{k_{j}} \tilde{u}_{n} \rangle
      + \langle D_{B^{a}} \tilde{u}_{n} | D_{k_{i}} \tilde{u}_{m} \rangle \langle \tilde{u}_{m} | D_{k_{j}} \tilde{u}_{n} \rangle
    )
    = \langle D_{B^{a}} \tilde{u}_{n} | \partial_{k_{i}} \left(
      \sum_{m} | \tilde{u}_{m} \rangle \langle \tilde{u}_{m} |
    \right) | D_{k_{j}} \tilde{u}_{n} \rangle
    = 0, \label{eq:identity}
  \end{equation}
  owing to the completeness of Bloch wave functions for each $\bm{k}$.

  \section{Spin Magnetic Octupole Moment} \label{app:spin_magnetic_octupole}
  In this Appendix, we rederive the spin magnetic octupole moment following Refs.~\cite{PhysRevB.112.134412,2504.21431}.
  The thermodynamic definition of the orbital or spin
  magnetic multipole moments~\cite{PhysRevB.97.134423,PhysRevB.98.020407,PhysRevB.98.060402,PhysRevB.99.024404,PhysRevB.112.134412,2504.21431}
  are based on the thermodynamic relation of the grand potential,
  \begin{equation}
    d \Omega
    = -S d T - N d \mu
    - (M_{a} - \partial_{x^{j}} M_{a}^{\phantom{a} j} + \partial_{x^{i}} \partial_{x^{j}} M_{a}^{\phantom{a} ij})
    B_{\mathrm{ext}}^{a}, \label{eq:thermodynamic_relation}
  \end{equation}
  where $S$ and $N$ are the entropy and particle number densities.
  The magnetic dipole $M_{a}$, quadrupole $M_{a}^{\phantom{a} j}$, and octupole moments $M_{a}^{\phantom{a} ij}$, are defined by
  \begin{subequations}\begin{align}
    M_{a}
    = & -\frac{\partial \Omega}{\partial B_{\mathrm{ext}}^{a}}, \label{eq:thermodynamic_dipole} \\
    M_{a}^{\phantom{a} j}
    = & -\frac{\partial \Omega}{\partial (\partial_{x^{j}} B_{\mathrm{ext}}^{a})}, \label{eq:thermodynamic_quadrupole} \\
    M_{a}^{\phantom{a} ij}
    = & -\frac{\partial \Omega}{\partial (\partial_{x^{i}} \partial_{x^{j}} B_{\mathrm{ext}}^{a})}. \label{eq:thermodynamic_octupole}
  \end{align}\label{eq:thermodynamic_multipole}\end{subequations}
  From these thermodynamic definitions, we obtain the Maxwell relations,
  \begin{subequations}\begin{align}
    \frac{\partial M_{a}}{\partial \mu}
    = & -\frac{\partial^{2} \Omega}{\partial \mu \partial B_{\mathrm{ext}}^{a}}
    = \frac{\partial N}{\partial B_{\mathrm{ext}}^{a}}, \label{eq:maxwell_dipole} \\
    \frac{\partial M_{a}^{\phantom{a} j}}{\partial \mu}
    = & -\frac{\partial^{2} \Omega}{\partial \mu \partial (\partial_{x^{j}} B_{\mathrm{ext}}^{a})}
    = \frac{\partial N}{\partial (\partial_{x^{j}} B_{\mathrm{ext}}^{a})}, \label{eq:maxwell_quadrupole} \\
    \frac{\partial M_{a}^{\phantom{a} ij}}{\partial \mu}
    = & -\frac{\partial^{2} \Omega}{\partial \mu \partial (\partial_{x^{i}} \partial_{x^{j}} B_{\mathrm{ext}}^{a})}
    = \frac{\partial N}{\partial (\partial_{x^{i}} \partial_{x^{j}} B_{\mathrm{ext}}^{a})}, \label{eq:maxwell_octupole}
  \end{align}\label{eq:maxwell}\end{subequations}
  Below, we compute the right-hand side of Eq.~\eqref{eq:maxwell} following Ref.~\cite{PhysRevB.112.134412}.

  The linear response of the particle number to a magnetic field
  $\langle \Delta \hat{n} \rangle(\Omega, \bm{Q})
  = \chi_{\hat{n} \hat{s}_{a}}^{\mathrm{R}}(\Omega, \bm{Q}) B_{\mathrm{ext}}^{a}(\Omega, \bm{Q})$
  is computed from the number-spin correlation function,
  \begin{equation}
    \chi_{\hat{n} \hat{s}_{a}}^{\mathrm{R}}(\Omega, \bm{Q})
    = -\sum_{nm} \int \frac{d^{d} k}{(2 \pi)^{d}}
    \langle u_{n}(\bm{k}_{-}) | u_{m}(\bm{k}_{+}) \rangle \langle u_{m}(\bm{k}_{+}) | \hat{s}_{a} | u_{n}(\bm{k}_{-}) \rangle
    \frac{
      f(\epsilon_{n}(\bm{k}_{-})) - f(\epsilon_{m}(\bm{k}_{+}))
    }{
      \hbar \Omega + \epsilon_{n}(\bm{k}_{-}) - \epsilon_{m}(\bm{k}_{+}) + i \eta
    }. \label{eq:number_spin}
  \end{equation}
  Here, we take the $\Omega \rightarrow 0$ limit first and expand up to the second order with respect to $\bm{Q}$.
  The distribution functions in Eq.~\eqref{eq:number_spin} are expanded as
  \begin{align}
    \frac{
      f(\epsilon_{n}(\bm{k}_{-})) - f(\epsilon_{n}(\bm{k}_{+}))
    }{
      \epsilon_{n}(\bm{k}_{-}) - \epsilon_{n}(\bm{k}_{+})
    }
    = & \frac{f(\epsilon_{n} - \delta_{n} + \Delta_{n}) - f(\epsilon_{n} + \delta_{n} + \Delta_{n})}{-2 \delta_{n}} \notag \\
    = & f^{\prime}(\epsilon_{n}) + \Delta_{n} f^{\prime \prime}(\epsilon_{n}) + \frac{1}{6} \delta_{n}^{2} f^{(3)}(\epsilon_{n}) \notag \\
    = & f^{\prime}(\epsilon_{n}) + \frac{Q_{i} Q_{j}}{24} [
      3 \partial_{k_{i}} \partial_{k_{j}} \epsilon_{n} f^{\prime \prime}(\epsilon_{n})
      + \partial_{k_{i}} \epsilon_{n} \partial_{k_{j}} \epsilon_{n} f^{(3)}(\epsilon_{n})
    ], \label{eq:static_intraband}
  \end{align}
  for $n = m$ and
  \begin{align}
    \frac{f(\epsilon_{n}(\bm{k}_{-})) - f(\epsilon_{m}(\bm{k}_{+}))}{\epsilon_{n}(\bm{k}_{-}) - \epsilon_{m}(\bm{k}_{+})}
    = & \frac{f(\epsilon_{n} - \delta_{n}) - f(\epsilon_{m} + \delta_{m})}{\epsilon_{n} - \epsilon_{m} - \delta_{n} - \delta_{m}} \notag \\
    = & \left(
      1 + \frac{\delta_{n} + \delta_{m}}{\epsilon_{n} - \epsilon_{m}}
    \right) \frac{
      f(\epsilon_{n}) - f(\epsilon_{m}) - \delta_{n} f^{\prime}(\epsilon_{n}) - \delta_{m} f^{\prime}(\epsilon_{m})
    }{
      \epsilon_{n} - \epsilon_{m}
    } \notag \\
    = & \frac{f(\epsilon_{n}) - f(\epsilon_{m})}{\epsilon_{n} - \epsilon_{m}}
    + \frac{Q_{j}}{2} \left[
      \frac{\partial_{k_{j}} (\epsilon_{n} + \epsilon_{m})}{\epsilon_{n} - \epsilon_{m}}
      \frac{f(\epsilon_{n}) - f(\epsilon_{m})}{\epsilon_{n} - \epsilon_{m}}
      - \frac{
        \partial_{k_{j}} \epsilon_{n} f^{\prime}(\epsilon_{n}) + \partial_{k_{j}} \epsilon_{m} f^{\prime}(\epsilon_{m})
      }{
        \epsilon_{n} - \epsilon_{m}
      }
    \right], \label{eq:static_interband}
  \end{align}
  for $n \not= m$, respectively.
  Note that the second-order terms with respect to $\bm{Q}$ are not necessary in Eq.~\eqref{eq:static_interband}
  because there is no interband matrix element of the zeroth order in Eq.~\eqref{eq:number_spin}, as seen below.

  Using the identities~\eqref{eq:matrix_element} and the spin magnetic quadrupole moment~\eqref{eq:spin_magnetic_quadrupole},
  the matrix elements in Eq.~\eqref{eq:number_spin} are expanded as
  \begin{align}
    & \langle u_{n}(\bm{k}_{-}) | u_{m}(\bm{k}_{+}) \rangle \langle u_{m}(\bm{k}_{+}) | \hat{s}_{a} | u_{n}(\bm{k}_{-}) \rangle \notag \\
    = & \tr [| u_{m}(\bm{k}_{+}) \rangle \langle u_{m}(\bm{k}_{+}) |
    \hat{s}_{a} | u_{n}(\bm{k}_{-}) \rangle \langle u_{n}(\bm{k}_{-})|] \notag \\
    = & \tr \left\{
      \left[
        | u_{m} \rangle \langle u_{m} | + \frac{Q_{j}}{2} \partial_{k_{j}} (| u_{m} \rangle \langle u_{m} |)
        + \frac{Q_{i} Q_{j}}{8} \partial_{k_{i}} \partial_{k_{j}} (| u_{m} \rangle \langle u_{m} |)
      \right]
    \right. \notag \\
    & \left.
      \times \hat{s}_{a} \left[
        | u_{n} \rangle \langle u_{n} | - \frac{Q_{j}}{2} \partial_{k_{j}} (| u_{n} \rangle \langle u_{n} |)
        + \frac{Q_{i} Q_{j}}{8} \partial_{k_{i}} \partial_{k_{j}} (| u_{n} \rangle \langle u_{n} |)
      \right]
    \right\} \notag \\
    = & \tr (| u_{m} \rangle \langle u_{m} | \hat{s}_{a} | u_{n} \rangle \langle u_{n} |)
    + \frac{Q_{j}}{2} \tr [
      \partial_{k_{j}} (| u_{m} \rangle \langle u_{m} |) \hat{s}_{a} | u_{n} \rangle \langle u_{n} |
      - | u_{m} \rangle \langle u_{m} | \hat{s}_{a} \partial_{k_{j}} (| u_{n} \rangle \langle u_{n} |)
    ] + \frac{Q_{i} Q_{j}}{8} \notag \\
    & \times \tr \{
      \partial_{k_{i}} \partial_{k_{j}} (| u_{m} \rangle \langle u_{m} |) \hat{s}_{a} | u_{n} \rangle \langle u_{n} |
      - [\partial_{k_{i}} (| u_{m} \rangle \langle u_{m} |) \hat{s}_{a} \partial_{k_{j}} (| u_{n} \rangle \langle u_{n} |)
      + (i \leftrightarrow j)]
      + | u_{m} \rangle \langle u_{m} | \hat{s}_{a} \partial_{k_{i}} \partial_{k_{j}} (| u_{n} \rangle \langle u_{n} |)
    \} \notag \\
    = & \tr (| u_{m} \rangle \langle u_{m} | \hat{s}_{a} | u_{n} \rangle \langle u_{n} |) \notag \\
    & + \frac{Q_{j}}{2} \tr [
      (| D_{k_{j}} u_{m} \rangle \langle u_{m} | + | u_{m} \rangle \langle D_{k_{j}} u_{m} |) \hat{s}_{a} | u_{n} \rangle \langle u_{n} |
      - | u_{m} \rangle \langle u_{m} | \hat{s}_{a} (| D_{k_{j}} u_{n} \rangle \langle u_{n} | + | u_{n} \rangle \langle D_{k_{j}} u_{n} |)
    ]
    + \frac{Q_{i} Q_{j}}{8} \notag \\
    & \times \tr \{
      \partial_{k_{i}} \partial_{k_{j}} (| u_{m} \rangle \langle u_{m} | \hat{s}_{a} | u_{n} \rangle \langle u_{n} |)
      - 2 [
        (| D_{k_{i}} u_{m} \rangle \langle u_{m} | + | u_{m} \rangle \langle D_{k_{i}} u_{m} |)
        \hat{s}_{a} (| D_{k_{j}} u_{n} \rangle \langle u_{n} | + | u_{n} \rangle \langle D_{k_{j}} u_{n} |) + (i \leftrightarrow j)
      ]
    \} \notag \\
    = & \langle u_{n} | \hat{s}_{a} | u_{n} \rangle \delta_{mn}
    + \frac{Q_{j}}{2} [
      -(\langle u_{n} | \hat{s}_{a} | D_{k_{j}} u_{n} \rangle - \langle D_{k_{j}} u_{n} | \hat{s}_{a} | u_{n} \rangle) \delta_{mn}
      - 2 \langle D_{k_{j}} u_{n} | u_{m} \rangle \langle u_{m} | \hat{s}_{a} | u_{n} \rangle
    ]
    + \frac{Q_{i} Q_{j}}{8} \notag \\
    & \times \{
      \partial_{k_{i}} \partial_{k_{j}} (\langle u_{n} | \hat{s}_{a} | u_{n} \rangle) \delta_{mn}
      - 2 [
        \langle D_{k_{i}} u_{n} | \hat{s}_{a} | D_{k_{j}} u_{n} \rangle \delta_{mn}
        + \langle D_{k_{i}} u_{n} | D_{k_{j}} u_{m} \rangle \langle u_{m} | \hat{s}_{a} | u_{n} \rangle \notag \\
        & - \langle D_{k_{i}} u_{n} | u_{m} \rangle
        (\langle u_{m} | \hat{s}_{a} | D_{k_{j}} u_{n} \rangle - \langle D_{k_{j}} u_{m} | \hat{s}_{a} | u_{n} \rangle)
        + (i \leftrightarrow j)]
    \} \notag \\
    \xleftarrow{\bm{B} \rightarrow 0} & \partial_{B^{a}} \tilde{\epsilon}_{n} \delta_{mn}
    + \frac{Q_{j}}{2} \{
      -[\langle D_{B^{a}} \tilde{u}_{n} | (\tilde{\epsilon}_{n} - \tilde{H}) | D_{k_{j}} \tilde{u}_{n} \rangle
      - \langle D_{k_{j}} \tilde{u}_{n} | (\tilde{\epsilon}_{n} - \tilde{H}) | D_{B^{a}} \tilde{u}_{n} \rangle] \delta_{mn} \notag \\
      & - 2 \langle D_{k_{j}} \tilde{u}_{n} | \tilde{u}_{m} \rangle [
        \partial_{B^{a}} \tilde{\epsilon}_{n} \delta_{mn}
        + (\tilde{\epsilon}_{n} - \tilde{\epsilon}_{m}) \langle \tilde{u}_{m} | D_{B^{a}} \tilde{u}_{n} \rangle
      ]
    \} \notag \\
    & + \frac{Q_{i} Q_{j}}{8} \left(
      \partial_{k_{i}} \partial_{k_{j}} \partial_{B^{a}} \tilde{\epsilon}_{n} \delta_{mn}
      - 2 \{
        \langle D_{k_{i}} \tilde{u}_{n} | \hat{s}_{a} | D_{k_{j}} \tilde{u}_{n} \rangle \delta_{mn}
        + \langle D_{k_{i}} \tilde{u}_{n} | D_{k_{j}} \tilde{u}_{m} \rangle [
          \partial_{B^{a}} \tilde{\epsilon}_{n} \delta_{mn} +
          (\tilde{\epsilon}_{n} - \tilde{\epsilon}_{m}) \langle \tilde{u}_{m} | D_{B^{a}} \tilde{u}_{n} \rangle
        ]
    \right. \notag \\
    & \left.
      - \langle D_{k_{i}} \tilde{u}_{n} | \tilde{u}_{m} \rangle [
        \partial_{B^{a}} (\tilde{\epsilon}_{n} + \tilde{\epsilon}_{m}) \langle \tilde{u}_{m} | D_{k_{j}} \tilde{u}_{n} \rangle
        + \langle D_{B^{a}} \tilde{u}_{m} | (\tilde{\epsilon}_{m} - \tilde{H}) | D_{k_{j}} \tilde{u}_{n} \rangle
        - \langle D_{k_{j}} \tilde{u}_{m} | (\tilde{\epsilon}_{n} - \tilde{H}) | D_{B^{a}} \tilde{u}_{n} \rangle
      ] + (i \leftrightarrow j)
      \}
    \right) \notag \\
    = & \partial_{B^{a}} \tilde{\epsilon}_{n} \delta_{mn}
    + \frac{Q_{j}}{2} [
      2 i \tilde{s}_{na}^{\phantom{na} j} \delta_{mn}
      - 2 (\tilde{\epsilon}_{n} - \tilde{\epsilon}_{m})
      \langle D_{k_{j}} \tilde{u}_{n} | \tilde{u}_{m} \rangle \langle \tilde{u}_{m} | D_{B^{a}} \tilde{u}_{n} \rangle
    ] \notag \\
    & + \frac{Q_{i} Q_{j}}{8} \left\{
      - \partial_{k_{i}} \partial_{k_{j}} \partial_{B^{a}} \tilde{\epsilon}_{n} \delta_{mn} / 3
      - 8 \tilde{s}_{na}^{\phantom{na} ij} \delta_{mn}
      - 2 [
        2 i \langle D_{k_{i}} \tilde{u}_{n} | \tilde{u}_{m} \rangle \tilde{s}_{mna}^{\phantom{mna} j}
        - \partial_{k_{j}} (\tilde{\epsilon}_{n} + \tilde{\epsilon}_{m})
        \langle D_{k_{i}} \tilde{u}_{n} | \tilde{u}_{m} \rangle \langle \tilde{u}_{m} | D_{B^{a}} \tilde{u}_{n} \rangle
    \right. \notag \\
    & \left.
        + (\tilde{\epsilon}_{n} - \tilde{\epsilon}_{m})
        (\langle D_{k_{i}} \tilde{u}_{n} | D_{k_{j}} \tilde{u}_{m} \rangle \langle \tilde{u}_{m} | D_{B^{a}} \tilde{u}_{n} \rangle
        + \langle D_{k_{i}} \tilde{u}_{n} | \tilde{u}_{m} \rangle \langle D_{k_{j}} \tilde{u}_{m} | D_{B^{a}} \tilde{u}_{n} \rangle)
        + (i \leftrightarrow j)
      ]
    \right\}. \label{eq:number_spin_2}
  \end{align}
  Here, we have introduced the intraband spin magnetic octupole moment as
  \begin{equation}
    \tilde{s}_{na}^{\phantom{na} ij}
    = \frac{1}{2} \langle D_{k_{i}} \tilde{u}_{n} |
    \frac{\partial_{B^{a}} \tilde{\epsilon}_{n} + \hat{s}_{a}}{2} | D_{k_{j}} \tilde{u}_{n} \rangle
    + \cc - \frac{1}{6} \partial_{k_{i}} \partial_{k_{j}} \partial_{B^{a}} \tilde{\epsilon}_{n}. \label{eq:spin_magnetic_octupole}
  \end{equation}
  This quantity is symmetric with respect to $i$ and $j$
  and reduced to the quantum metric if $\hat{s}_{a}$ is replaced by $1$.
  As mentioned above, there is no interband matrix element in the zeroth-order term.

  Finally, we combine Eq.~\eqref{eq:number_spin_2} with Eqs.~\eqref{eq:static_intraband} and \eqref{eq:static_interband}.
  The zeroth-order coefficients with respect to $\bm{Q}$ originate from $n = m$ only,
  \begin{equation}
    \frac{\partial S_{a}}{\partial \mu}
    \xleftarrow{\bm{B} \rightarrow 0} -\sum_{n} \int \frac{d^{d} k}{(2 \pi)^{d}}
    \partial_{B^{a}} \tilde{\epsilon}_{n} f^{\prime}(\tilde{\epsilon}_{n}), \label{eq:maxwell_dipole_2}
  \end{equation}
  leading to the spin expectation value,
  \begin{equation}
    S_{a}
    \xleftarrow{\bm{B} \rightarrow 0} \sum_{n} \int \frac{d^{d} k}{(2 \pi)^{d}}
    \partial_{B^{a}} \tilde{\epsilon}_{n} f(\tilde{\epsilon}_{n}). \label{eq:thermodynamic_dipole_2}
  \end{equation}
  Hereafter, we denote the spin magnetic multipole moments as $S_{a}^{\dots}$ instead of $M_{a}^{\dots}$.
  The first-order coefficients with respect to $\bm{Q}$ come from $n = m$ and $n \not= m$,
  \begin{align}
    \frac{\partial S_{a}^{\phantom{a} j}}{\partial \mu}
    \xleftarrow{\bm{B} \rightarrow 0} & -\sum_{n} \int \frac{d^{d} k}{(2 \pi)^{d}} \left\{
      \tilde{s}_{na}^{\phantom{na} j} f^{\prime}(\tilde{\epsilon}_{n})
      + \sum_{m (\not= n)}
      i \langle D_{k_{j}} \tilde{u}_{n} | \tilde{u}_{m} \rangle \langle \tilde{u}_{m} | D_{B^{a}} \tilde{u}_{n} \rangle
      [f(\tilde{\epsilon}_{n}) - f(\tilde{\epsilon}_{m})]
    \right\} \notag \\
    = & -\sum_{n} \int \frac{d^{d} k}{(2 \pi)^{d}} \left[
      \tilde{s}_{na}^{\phantom{na} j} f^{\prime}(\tilde{\epsilon}_{n})
      + \sum_{m (\not= n)}
      (i \langle D_{k_{j}} \tilde{u}_{n} | \tilde{u}_{m} \rangle \langle \tilde{u}_{m} | D_{B^{a}} \tilde{u}_{n} \rangle + \cc)
      f(\tilde{\epsilon}_{n})
    \right], \label{eq:maxwell_quadrupole_2}
  \end{align}
  which leads to the spin magnetic quadrupole moment~\cite{PhysRevB.97.134423,PhysRevB.99.024404},
  \begin{align}
    S_{a}^{\phantom{a} j}
    \xleftarrow{\bm{B} \rightarrow 0} & \sum_{n} \int \frac{d^{d} k}{(2 \pi)^{d}} \left[
      \tilde{s}_{na}^{\phantom{na} j} f(\tilde{\epsilon}_{n})
      + \sum_{m (\not= n)}
      (i \langle D_{k_{j}} \tilde{u}_{n} | \tilde{u}_{m} \rangle \langle \tilde{u}_{m} | D_{B^{a}} \tilde{u}_{n} \rangle + \cc)
      f^{(-1)}(\tilde{\epsilon}_{n})
    \right]. \label{eq:thermodynamic_quadrupole_2}
  \end{align}
  Using Eq.~\eqref{eq:spin_magnetic_quadrupole_nonhermitian} and the identity~\eqref{eq:identity},
  the second-order coefficients with respect to $\bm{Q}$ are expressed by
  \begin{align}
    \frac{\partial S_{a}^{\phantom{a} ij}}{\partial \mu}
    \xleftarrow{\bm{B} \rightarrow 0} & -\frac{1}{2} \sum_{n} \int \frac{d^{d} k}{(2 \pi)^{d}} \left(
      \left(
        \frac{1}{24} \partial_{k_{i}} \partial_{k_{j}} \partial_{B^{a}} \tilde{\epsilon}_{n} + \tilde{s}_{na}^{\phantom{na} ij}
      \right)
      f^{\prime}(\tilde{\epsilon}_{n})
      - \frac{1}{24} \partial_{B^{a}} \tilde{\epsilon}_{n}
      [3 \partial_{k_{i}} \partial_{k_{j}} \tilde{\epsilon}_{n} f^{\prime \prime}(\tilde{\epsilon}_{n})
      + \partial_{k_{i}} \tilde{\epsilon}_{n} \partial_{k_{j}} \tilde{\epsilon}_{n} f^{(3)}(\tilde{\epsilon}_{n})]
    \right. \notag \\
      & + \sum_{m (\not= n)} \left\{
        \left[
          \frac{
            i \langle D_{k_{i}} \tilde{u}_{n} | \tilde{u}_{m} \rangle \tilde{s}_{mna}^{\phantom{mna} j}
          }{
            \tilde{\epsilon}_{n} - \tilde{\epsilon}_{m}
          }
          + \frac{1}{2} (
            \langle D_{k_{i}} \tilde{u}_{n} | D_{k_{j}} \tilde{u}_{m} \rangle \langle \tilde{u}_{m} | D_{B^{a}} \tilde{u}_{n} \rangle
            + \langle D_{k_{i}} \tilde{u}_{n} | \tilde{u}_{m} \rangle \langle D_{k_{j}} \tilde{u}_{m} | D_{B^{a}} \tilde{u}_{n} \rangle
          )
        \right]
      \right. \notag \\
    & \left.
      \left.
        \times [f(\tilde{\epsilon}_{n}) - f(\tilde{\epsilon}_{m})]
        - \frac{1}{2} \langle D_{k_{i}} \tilde{u}_{n} | \tilde{u}_{m} \rangle \langle \tilde{u}_{m} | D_{B^{a}} \tilde{u}_{n} \rangle [
          \partial_{k_{j}} \tilde{\epsilon}_{n} f^{\prime}(\tilde{\epsilon}_{n})
          + \partial_{k_{j}} \tilde{\epsilon}_{m} f^{\prime}(\tilde{\epsilon}_{m})
        ]
      \right\}
    \right) + (i \leftrightarrow j) \notag \\
    = & -\frac{1}{2} \sum_{n} \int \frac{d^{d} k}{(2 \pi)^{d}} \left\{
      \tilde{s}_{na}^{\phantom{na} ij} f^{\prime}(\tilde{\epsilon}_{n})
      - \frac{1}{12} \partial_{B^{a}} \tilde{\epsilon}_{n} \partial_{k_{i}} \partial_{k_{j}} \tilde{\epsilon}_{n}
      f^{\prime \prime}(\tilde{\epsilon}_{n})
    \right. \notag \\
    & \left.
      + \sum_{m (\not= n)} \left[
        \frac{
          i \langle D_{k_{i}} \tilde{u}_{n} | \tilde{u}_{m} \rangle \tilde{s}_{mna}^{\phantom{mna} j} + \cc
        }{
          \tilde{\epsilon}_{n} - \tilde{\epsilon}_{m}
        } f(\tilde{\epsilon}_{n})
        - \frac{1}{2}
        (\langle D_{k_{i}} \tilde{u}_{n} | \tilde{u}_{m} \rangle \langle \tilde{u}_{m} | D_{B^{a}} \tilde{u}_{n} \rangle + \cc)
        \partial_{k_{j}} \tilde{\epsilon}_{n} f^{\prime}(\tilde{\epsilon}_{n})
      \right]
    \right\} + (i \leftrightarrow j), \label{eq:maxwell_octupole_2}
  \end{align}
  leading to the spin magnetic octupole moment~\cite{PhysRevB.112.134412,2504.21431},
  \begin{align}
    S_{a}^{\phantom{a} ij}
    = & \frac{1}{2} \sum_{n} \int \frac{d^{d} k}{(2 \pi)^{d}} \left\{
      \tilde{s}_{na}^{\phantom{na} ij} f(\tilde{\epsilon}_{n})
      - \frac{1}{12} \partial_{B^{a}} \tilde{\epsilon}_{n} \partial_{k_{i}} \partial_{k_{j}} \tilde{\epsilon}_{n}
      f^{\prime}(\tilde{\epsilon}_{n})
    \right. \notag \\
    & \left.
      + \sum_{m (\not= n)} \left[
        \frac{
          i \langle D_{k_{i}} \tilde{u}_{n} | \tilde{u}_{m} \rangle \tilde{s}_{mna}^{\phantom{mna} j} + \cc
        }{
          \tilde{\epsilon}_{n} - \tilde{\epsilon}_{m}
        } f^{(-1)}(\tilde{\epsilon}_{n})
        - \frac{1}{2}
        (\langle D_{k_{i}} \tilde{u}_{n} | \tilde{u}_{m} \rangle \langle \tilde{u}_{m} | D_{B^{a}} \tilde{u}_{n} \rangle + \cc)
        \partial_{k_{j}} \tilde{\epsilon}_{n} f(\tilde{\epsilon}_{n})
      \right]
    \right\} + (i \leftrightarrow j). \label{eq:thermodynamic_octupole_2}
  \end{align}
  This quantity is by construction symmetric with respect to $i$ and $j$
  and reduced to the thermodynamic electric quadrupole moment~\cite{PhysRevB.102.235149} if $\hat{s}_{a}$ is replaced by $1$.
  Comparing Eq.~\eqref{eq:maxwell_octupole_2} with Eq.~\eqref{eq:spin_accumulation_2},
  we obtain $(g_{sa}^{(\mathrm{II}) ij} + g_{sa}^{(\mathrm{II}) ji}) / 2 = q (\partial S_{a}^{\phantom{a} ij} / \partial \mu)$
  as far as the system is insulating.
  \end{widetext}
\end{document}